# Forward and inverse mantle convection with neural operators

Chenxi Kong, Michael Gurnis, Zachary E Ross

# Forward and inverse mantle convection with neural operators

Chenxi Kong, Michael Gurnis and Zachary E. Ross

*Seismological Laboratory, California Institute of Technology, Pasadena, CA 91125, USA. E-mail: ckong@caltech.edu*

## SUMMARY

Thermal state reconstruction—reversing convection to recover the thermal structure of the mantle at an earlier geologic time—is an important tool to understand the evolution of mantle convection and its relation to seismic tomographic images and observations at the surface. Thermal state reconstructions are computationally expensive. Here we transformed the basic computational element, numerical solvers, into neural operators, a class of machine learning models for learning mappings between function spaces. Focusing on a specific architecture, Fourier Neural Operators, we demonstrate that they can represent not only a surrogate model like the Stokes system of equations using a purely physics informed approach, but also discover operators without explicit mathematical formulations or even ill-posedness from data, including the direct mapping between two convecting thermal states separated by a long time interval much larger than the Courant–Friedrichs–Lewy condition and its reversal. These neural operators significantly accelerate forward and inverse convection modelling by transforming forward physical processes into surrogate models with lower complexity while utilizing auto-differentiation to calculate gradients. With this framework, we demonstrate the strengths and weaknesses of four methods for thermal state reconstructions: reverse buoyancy, reverse convection operator, an inversion with only the terminal thermal state, and a joint inversion with the terminal thermal state and surface velocity evolution. The reverse convection operator is shown to perform poorly in the presence of observational noise, but the joint inversion overcomes this limitation. The joint technique could probably become a solution to large-scale thermal state inversion problems using seismic tomography and plate tectonic reconstructions.

## 1 INTRODUCTION

Plate kinematics, the subsidence and cooling of oceanic plates and the life cycle of plates demonstrate that plates make up the upper thermal boundary layer of mantle convection (G.F. Davies 1999). Central to this concept is that plate motions reflect the surface, horizontal velocity field of convection. With advances in seismic tomography, a more detailed picture of the present day mantle, and hence the temperature field of mantle convection, has gradually emerged (e.g. J. Ritsema *et al.* 2011; P. Koelemeijer *et al.* 2016; C. Lu *et al.* 2019). If we can construct a self-consistent dynamic framework of the present Earth (the dynamics that links the driving temperature field with the surface kinematics), the possibility opens that we can build ones for earlier times, especially in light of the extensive constraints that exist for past plate motions (e.g. M. Seton *et al.* 2012; A.S. Merdith *et al.* 2021).

Such a framework could be based on a forward model describing the mantle dynamics. Forward models have successfully reproduced many features of plate tectonics, such as ones that model global plate motions (e.g. B.H. Hager & R.J. O'Connell 1981; C.P. Conrad & C. Lithgow-Bertelloni 2002; G. Stadler *et al.* 2010; J. Hu *et al.* 2024), those that attempt to simulate plate tectonics (e.g. H. Van Heck & P. Tackley 2008; T.W. Becker & L. Fuchs 2023), or those that produce realistic subduction (e.g. S. Zhong & M. Gurnis 1995; M.I. Billen 2008; S. Goes *et al.* 2017). Such later plate tectonic and subduction models are initiated from a specified state in parameter space and produce evolutionary pathways. The models which best fit a set of observations, such as plate kinematics and present-day mantle structure, then partially reveal the physics underlying the tectonic process. However, there exist substantial uncertainties within a large parameter space and finding the optimal model is often difficult.

Beside forward modelling trials, other approaches have been explored, including those aimed at reconstructing a previous mantle state, or reversing mantle convection (e.g. H.P. Bunge *et al.* 2003; C.P. Conrad & M. Gurnis 2003; A. Ismail-Zadeh *et al.* 2004; L. Liu & M. Gurnis 2008; D. Li *et al.* 2017). As mantle convection is dissipative and irreversible, reconstructing prior states is an ill-posed problem and cannot be solved directly. Consequently, the reconstruction problem is usually solved through an inverse approach, overcoming the ill-posedness, while allowing additional information to be used as constraints (A. Kirsch *et al.* 2011). As alluded to, there exist rich chronological data preserved at Earth's surface providing records of surface imprints of interior dynamics. Those data are taken in various forms including but not limited to: Plate kinematics inferred from seafloor spreading, paleomagnetic reconstruction of continental motions and hotspot tracks (e.g. D.C. Engebretson *et al.* 1985; B. Steinberger & R.J. O'Connell 1998; T.H. Torsvik *et al.* 2008;

M. Seton *et al.* 2012); topography revealed by stratigraphic sequences and past sea level changes (e.g. M. Gurnis 1993; S. Spasojevic *et al.* 2009); mantle flow history recorded within mineral anisotropy (e.g. N.M. Ribe 1989; M.D. Long & T.W. Becker 2010); and subduction and intraplate deformations preserved in exhumed and magmatic rock sequences (e.g. N.B. Harris *et al.* 1986; W. Ernst 1988). Besides the chronological data, the terminal state—the observed present-day mantle—provides direct constraint across the entire mantle depth. The observables are connected to the system state variable, the thermal structure of the mantle, through a fully dynamic and self-consistent forward model. With an objective function, mismatch between model predictions and observables, the update direction of the initial state can be calculated by solving the adjoint state equations and optimized towards that initial state by iterative gradient descent (D. Li *et al.* 2017) .

Although a gradient descent method has guided optimization directions that avoids empirical trials, it still relies on an iterative approach. Consequently, inversions have a high computational cost which can be several orders of magnitude larger than one forward model (D. Li *et al.* 2017). Solving the forward and adjoint convection equations iteratively with traditional partial differential equation (PDE) solvers is computationally demanding and has limited its applicability for geophysical inversions. Leaving aside the iterations, solving the forward model itself (and its adjoint) can be expensive as well. To model realistic mantle dynamics with plate tectonics requires both a long temporal scale and a fine spatial resolution at plate boundaries, as plates recycle (M.I. Billen 2008; J. Hu *et al.* 2024) over hundreds of million years (N. Zhang *et al.* 2010). A global reconstruction that incorporates realistic, evolving plate tectonics, is computationally intractable with traditional numerical methods.

We attempt to overcome this challenge, by reformulating the governing equations of thermal convection into a surrogate model with a neural operator (N. Kovachki *et al.* 2023). With the rapid development in machine learning, there have been studies using deep learning approaches in mantle dynamics modelling and inversions. S. Atkins *et al.* (2016) and M. Shahnas *et al.* (2018) used neural networks to infer the controlling parameters in convection from synthesized thermal fields, but not within an inverse approach. Later, S. Agarwal *et al.* (2021, 2025) formulated a convolutional neural network (CNN; Y. LeCun *et al.* 1998) from computations of thermal convection. The surrogate models show an appealing speedup in forward modelling, but their accuracy in tracking long-term unsteady convection is insufficient to capture the motion of plume structures, and they have not yet been applied to inverse problems. Neural operators inherit almost all of the advantages of a common neural network, including fast computations (evaluation) once trained for both forward and backward solves. Architecturally, it is designed to learn the integral kernel of mappings instead of point-wise connections between nodal values; the inputs and outputs are not forced to be discretized on a specific mesh and the neural operators learn the mappings between function spaces. The most common scenario is to learn a given PDE operator. Neural operators can also learn mappings that cannot normally be explicitly described mathematically or numerically solved, collectively known as operator discovery. Consequently, neural operators have been widely used in scientific computations (K. Azizzadenesheli *et al.* 2024), including for fluid dynamics applications to turbulent flow (Navier–Stokes), Darcy flow, Stokes flow and Rayleigh–Bénard convection (Z. Li *et al.* 2021; T. Marwah *et al.* 2023; M. Straat *et al.* 2025). In geophysics, examples include learning the operator for seismic wave propagation and its incorporation into full waveform inversion (Y. Yang *et al.* 2021; C. Zou *et al.* 2024, 2025). Various neural operator architectures are emerging, among which the Fourier neural operator (FNO), which parametrizes the kernel in Fourier space, has been extensively used and has been shown to be robust for high accuracy operator learning (Z. Li *et al.* 2021). Here, we use FNO as the basic neural operator architecture. We test various possibilities for the usage of neural operators in modelling forward and reverse mantle convection. We find that this surrogate approach can significantly accelerate both forward and inverse computation, while keeping a total workflow cost (including training) comparable to that of performing a time-dependent inversion only once by a traditional numerical method.

## 2 METHODOLOGY

### 2.1 Thermal convection problem

The mantle is treated as an incompressible, creeping viscous fluid. Assuming chemical homogeneity, it is simplified to basally heated thermal convection within a 2-D Cartesian domain $\Omega$, governed by the conservation of mass, momentum and energy (S. Zhong *et al.* 2000)

$$\nabla \cdot \boldsymbol{u} = 0, \tag{1}$$

$$-\nabla p + \nabla \cdot \left[\eta \left(\nabla \boldsymbol{u} + (\nabla \boldsymbol{u})^{\top}\right)\right] + \alpha \rho_0 (T - T_0) \boldsymbol{e}_z = \boldsymbol{0}, \tag{2}$$

$$\frac{\partial T}{\partial t} + \boldsymbol{u} \cdot \nabla T - \nabla \cdot (\kappa \nabla T) = 0, \tag{3}$$

where $\boldsymbol{u}$, $p$, $\eta$, $\alpha$, $\rho_0$, $T$, $T_0$ and $\kappa$ are velocity, dynamic pressure, dynamic viscosity, thermal expansitivity, reference density, temperature, reference temperature and thermal diffusivity, respectively. $\eta$, $\alpha$ and $\kappa$ are set as constant values. By normalizing with $\eta = \eta_0$, $\alpha = \alpha_0$, $\kappa = \kappa_0$, $\boldsymbol{x} = D_0 \boldsymbol{x}'$, $p = \frac{\eta_0 \kappa_0}{D_0^2} p'$, $\boldsymbol{u} = \frac{\kappa_0}{D_0} \boldsymbol{u}'$ and $T = \Delta T T' + T_0$, and dropping the primes subsequently, we have the dimensionless governing equations,

$$\nabla \cdot \boldsymbol{u} = 0, \tag{4}$$

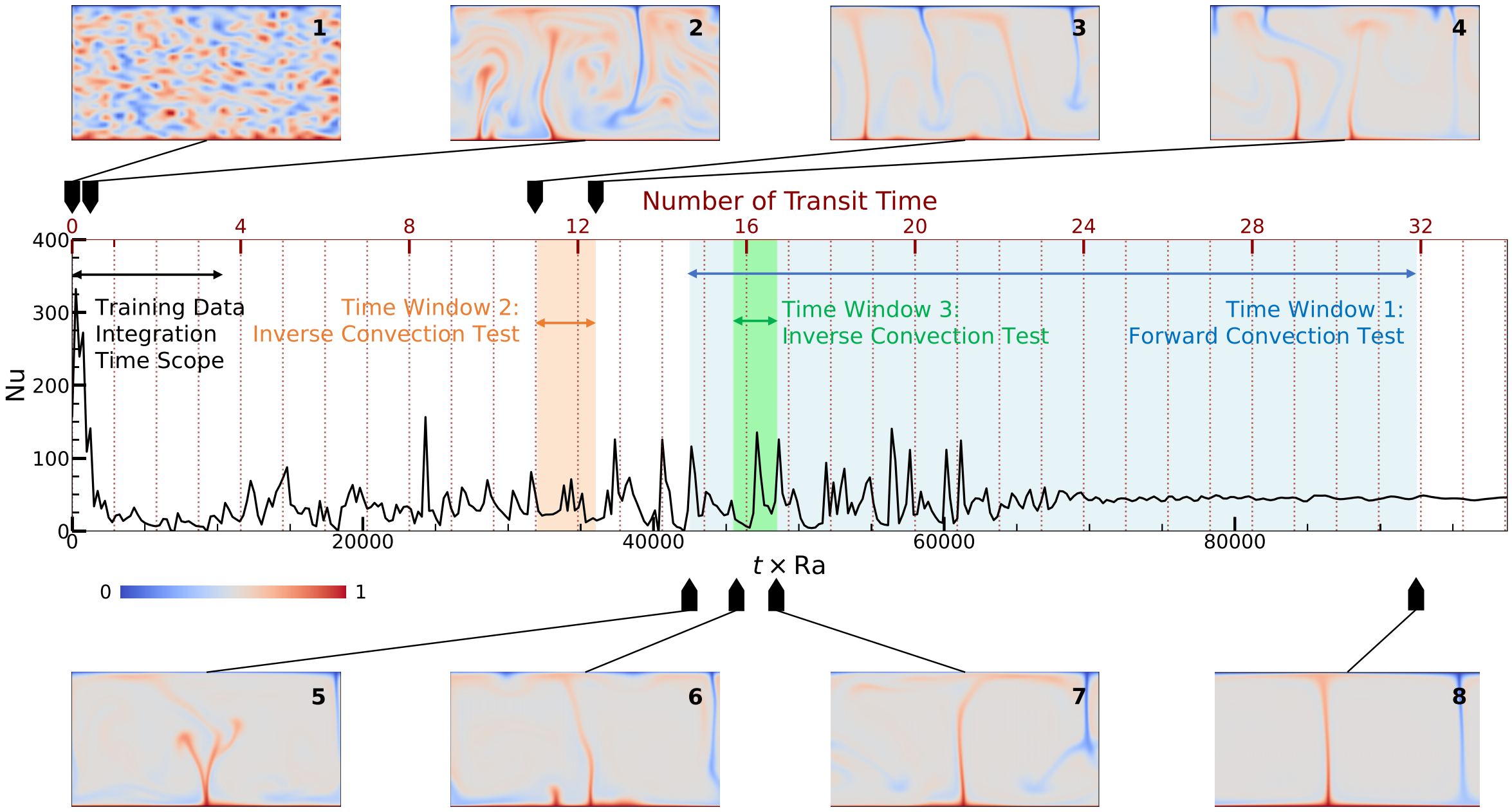

**Figure 1.** An example of forward convection computed with `Underworld`, which is used as an evaluation data sequence in this study. The computation is initiated from a Gaussian random initial thermal field that satisfies the prescribed boundary conditions. The total integration time is around 34 transit times, during which the convection pattern evolves from a chaotic one to a relatively steady state, as shown by the tracked Nusselt number. The temperature field is displayed at eight instances to depict the thermal evolution, with values shown over the range [0, 1]. The same colourmap and normalization are used for temperature throughout the paper.

$$-\nabla p + \nabla^2 \boldsymbol{u} + \mathrm{Ra} T \boldsymbol{e}_z = \boldsymbol{0}, \tag{5}$$

$$\frac{\partial T}{\partial t} + \boldsymbol{u} \cdot \nabla T - \nabla^2 T = 0, \tag{6}$$

where $\mathrm{Ra} = \frac{\rho g \alpha_0 \Delta T D_0^3}{\eta_0 \kappa_0}$ is the Rayleigh number. For the unknowns, we introduce their solution spaces defined on $\Omega = [0, L_x] \times [0, D]$, that $\boldsymbol{u} \in \mathcal{U}$, $p \in \mathcal{P}$, $T \in \mathcal{T}$. The governing equations can be decomposed into two sets: the Stokes (eqs 4 and 5) and the advection–diffusion (eq. 6). The Stokes equations solve velocities from the buoyancy field determined by temperature: $\boldsymbol{S} : \mathcal{T} \mapsto \mathcal{U}$. The advection–diffusion equation integrates advection $\boldsymbol{A} : (\mathcal{T}, \mathcal{U}) \mapsto \mathcal{T}$ and diffusion $\boldsymbol{D} : \mathcal{T} \mapsto \mathcal{T}$ on $T$ for a time step, $\Delta t$, where $\boldsymbol{S}$, $\boldsymbol{A}$, $\boldsymbol{D}$ denote their solution operators. The solutions to both equations are subject to the boundary conditions. For the Stokes, we assume free-slip conditions on the top and bottom and for the advection–diffusion equation, we assume a Dirichlet boundary condition for temperature, where $T = 0$ at the top and $T = 1$ at the bottom. For the side walls, periodic boundary conditions are assumed.

Traditionally, the solution of the two components is interwoven and the dynamic process of thermal convection emerges. Numerical methods, such as the finite element method (FEM; e.g. U. Christensen 1984) or finite differences (FD; e.g. T. Gerya 2019), have been applied for the solves. Powerful software tools have emerged for the study of geodynamic problems in recent decades and we use an open source, well-benchmarked FEM-based program, `Underworld`, as the numerical solver for creating sample and validation data (J. Mansour *et al.* 2020). An example computation from `Underworld` is given in Fig. 1, where the Rayleigh number is $10^7$ with the domain discretized with $256 \times 256$ linear elements; the numerical procedure used will be detailed below. The computation is initiated with a random Gaussian thermal field that satisfies the boundary conditions, and is integrated forward for about 34 transit times, $t_{\mathrm{tr}}$, a characteristic time defined as that needed for a point to traverse the box depth. For the Earth, the transit time is ~50 Myr (S. Zhong *et al.* 2007) (see Supplementary Material for computing method used in this study). With traditional solvers, the time step size of each forward integration is limited by the Courant–Friedrichs–Lewy (CFL) condition (R. Courant *et al.* 1928). In the following discussion, we will show that neural operators can circumvent this limit, greatly accelerating computations. Assuming a constant, maximum $\Delta t$ that could satisfy the limit throughout the evolving sequence, the computation can be described as applying an autoregressive operator $\boldsymbol{F}^{+\Delta t} : \mathcal{T} \mapsto \mathcal{T}$ recursively on a state variable of $T$, by which we define $\boldsymbol{F}^{+\Delta t}$ as:

$$\begin{aligned} \boldsymbol{F}^{+\Delta t}(\cdot) :&= \boldsymbol{I}(\cdot) + \Delta t(\boldsymbol{A}(\cdot, \boldsymbol{u}) + \boldsymbol{D}(\cdot)) \\ &= \boldsymbol{I}(\cdot) + \Delta t(\boldsymbol{A}(\cdot, \boldsymbol{S}(\cdot)) + \boldsymbol{D}(\cdot)) \end{aligned} \tag{7}$$

with $\boldsymbol{F}^{+n\Delta t} := \left(\boldsymbol{F}^{+\Delta t}\right)^n$ is the mapping from an initial thermal state to a future one at $n\Delta t$.

### 2.2 Thermal state reconstruction

In addition to the forward solution of mantle convection, we are concerned with its evolution through reconstructing previous states from observations. One must first address whether a previous thermal state can be reconstructed as thermal convection is dissipative: New information is generated with amplitudes magnified at thermal boundary layers, but then vanishes into noise through advection and diffusion. With the diffusion time scaling in proportion to the square of size, the finest structures diffuse most rapidly, while the larger has a longer, persistent influence, perhaps extending over a substantial period of geologic time. Within a finite duration, some dominant, longer wavelength features will likely still be recoverable from the remnants, although lacking reliable high-frequency details. This is the basis that supports the feasibility of mantle state reconstruction.

A potential avenue for reconstructing a previous thermal state is to temporally reverse convection by replacing the forward time $t$ in eqs (4)–(6) with $-s$, which is the reversal time.

$$\nabla \cdot \boldsymbol{u} = 0, \tag{8}$$

$$-\nabla p + \nabla^2 \boldsymbol{u} + \mathrm{Ra} T \boldsymbol{e}_z = \boldsymbol{0}, \tag{9}$$

$$\frac{\partial T}{\partial s} - \boldsymbol{u} \cdot \nabla T + \nabla^2 T = 0. \tag{10}$$

Similar to the forward problem, eqs (8)–(10) form an operator $\boldsymbol{F}^{-n\Delta t}$, that maps a thermal field to its past state at $-n\Delta t$. However, $\boldsymbol{F}^{-n\Delta t}$ is ill-posed. The forward dynamics forms a dissipative system that tends to compress the state trajectory into an attractor (Fig. 1). In contrast, the reverse dynamics system is expansive in which any noise is magnified rapidly because of the existence of the anti-diffusion operator, the negative Laplacian. If one attempts to solve eqs (8)–(10) with numerical methods directly, the amplified error in the thermal field grows rapidly within several CFL time steps. Hence, a simple but inaccurate approach to reversing convection is to retain the diffusion process while integrating backwards (C.P. Conrad & M. Gurnis 2003), where the advection diffusion equation reads

$$\frac{\partial T}{\partial s} - \boldsymbol{u} \cdot \nabla T - \nabla^2 T = 0, \tag{11}$$

which only reverses the direction of advection, or the sign of the buoyancy force. We will call this the 'reverse buoyancy method'.

As detailed subsequently, important improvements for thermal state reconstructions can be made with neural operators. We will approximate the ill-posed operator $\boldsymbol{F}^{-n\Delta t}$ directly with a neural surrogate $\boldsymbol{F}_{\phi}^{-n\Delta t}$ capable of suppressing the rapid magnification of numerical noise. That is, we will train a neural operator that aims to make the most informative prediction directly towards a previous state by the limit of diffusion.

Nevertheless, the approximated reverse convection neural operator is still sensitive to the noise within the input, making its application in geophysical problems limited and difficult. Hence, the reconstruction problem has been commonly regarded as an inverse problem covering a temporal scope $[-N \times n\Delta t, 0]$ with the target being a past mantle state $T_0 := T(t = -N \times n\Delta t)$ (H.P. Bunge *et al.* 2003; A. Ismail-Zadeh *et al.* 2004; L. Liu & M. Gurnis 2008; D. Li *et al.* 2017). Inverse methods are more robust against observational noise. Similar to the direct methods mentioned above where only the observation of the terminal thermal state is utilized, the inversion can be formulated simply by a fit to the terminal state. However, the constraint can also take in various forms of observational data beyond that; this is important with the rich chronological records constraining mantle evolution. We will later demonstrate that compared with the inversion using the terminal thermal state only, incorporation of chronological data could greatly improve the accuracy of inversion and enlarge the time span for reconstruction. By contrast, this approach is named 'joint inversion', where two classes of observations are used. First, information on present-day mantle structure defined in $\Omega$, often revealed by seismic tomography (or other geophysical imaging); second, chronological data defined along the time axis on the top boundary $L_s$, such as surface horizontal velocity records (corresponding to plate kinematics) $\boldsymbol{v}_x^{\mathrm{obs}}(t)$, $t \in [-N \times n\Delta t, 0]$. For simplicity, we assume that the terminal thermal state $T_{\mathrm{N}} := T(t = 0)$, and $\boldsymbol{v}_x^{\mathrm{obs}}(t)$ are known. The two constraints are applied through the governing equations by connecting the inversion target to the observations.

The inversion is conducted through an iterative approach. We will use superscript $k$ to denote the iteration step index. When $k = 0$, one would need to start from an initial guess to the target state $T_0^0$, and compute the forward evolution process to determine the values of diagnostic variables, which correspond to observations, that is, the surface velocity of each time step $\boldsymbol{v}_{x,i}^0$, $i = 0, \cdots, N$, and the mantle thermal field at $t = N \times n\Delta t$, defined as $T_n^0$. Then, by comparing the computed diagnostic variables to observations, we calculate the objective function (Modified from D. Li *et al.* 2017)

$$\mathcal{J}^k = \frac{\beta_1}{|\Omega|} \int_{\Omega} (T_{\mathrm{N}}^k - T_{\mathrm{N}}^{\mathrm{obs}})^2 \mathrm{d}w + \frac{\beta_2}{n|L_{\mathrm{s}}|} \sum_{i=0}^{n} \int_{L_{\mathrm{s}}} (\boldsymbol{v}_i^k - \boldsymbol{v}_i^{\mathrm{obs}})^2 \mathrm{d}l + \mathcal{P}(T_0^k) \quad , \tag{12}$$

where $|\cdot|$ denotes the Lebesgue measure of a subdomain in which the diagnostic variables are evaluated. In this study with a 2-D Cartesian geometry, $|\Omega|$ defines the area of computational domain and $|L_{\mathrm{s}}|$ defines the length of the upper boundary. The first and second terms evaluate the misfits between the model predictions and observations of terminal state and surface horizontal velocities. The regularization function $\mathcal{P}$ combines Laplacian smoothing and a penalty on departures of the initial state from its spatial mean

$T_{\text{mean}}$:

$$\mathcal{P}(T_0^k) = \frac{1}{|\Omega|}\int_\Omega \left(\beta_3(\nabla^2 T_0^k)^2 + \beta_4(T_0^k - T_{\text{mean}})^2\right) \mathrm{d}w \tag{13}$$

$\beta_1$ to $\beta_4$ are weighting coefficients of each term (Table S1, Supplementary Material). Then, a gradient of the objective function with respect to the target state, $\nabla_{T_0^k}\mathcal{J}^k$, from which the target state can be updated through gradient descent by the Adam algorithm (D.P. Kingma & J. Ba 2015):

$$T_0^{k+1} = T_0^k - \alpha_k \cdot \text{Adam}\left(\nabla_{T_0^k}\mathcal{J}^k\right). \tag{14}$$

In traditional numerical methods, $\nabla_{T_0^k}\mathcal{J}^k$ is calculated by solving the adjoint equation, which shares a similar form as the forward computation. The total computational cost would be proportional to the product of time steps and the iteration steps, making the inversion typically several orders of magnitude more demanding than the forward problem (e.g. D. Li *et al.* 2017). In cases where the solution to $\boldsymbol{F}^{+n\Delta t}$ is already expensive, the mantle state reconstruction would then be computationally prohibitive. Hence, this motivates the use of neural operators to replace the numerical solvers in thermal reconstruction problems.

In all, four different reconstruction methods are formulated in this study, and their performances are compared against each other: (1) reverse buoyancy, (2) reverse convection neural operator, (3) inversion with forward convection neural operator using the terminal state only and (4) joint inversion with forward convection neural operator.

### 2.3 Neural operators

Given geodynamic forward models and mantle state reconstructions (consisting of repeated computation of $\boldsymbol{F}^{+\Delta t}$), the workflow would be significantly accelerated if this basic operation could be replaced with a faster and less costly approach. This is the inspiration for a new deep learning workflow as a replacement for the PDE-formulated operator, utilizing rapidly developing technology in machine learning and graphics processing unit (GPU) computing. The approach transforms the problem of solving PDEs into a forward pass through a deep learning model, while computing derivatives on a clearly structured, net-shaped computational graph with fast auto-differentiation (A. Griewank & A. Walther 2008). Neural operators are a class of models that generalize neural networks to function spaces and are universal approximators of nonlinear operators (N. Kovachki *et al.* 2023). In general, it approximates a functional mapping $\mathcal{G} : \mathcal{A} \mapsto \mathcal{U}$ with the following architecture $\mathcal{G}_\phi$, parametrized by the trainable $\phi$:

$$\begin{aligned} &h_0(x) = P\left(a(x)\right) \\ &h_{i+1}(x) = \sigma\left(Wh_i(x) + \int \kappa_\phi(x, y, h_i(x), h_i(y))h_i(y)dy + b(x)\right) \quad i = 0, 1, \cdots, D-1. \\ &u(x) = Q\left(h_D(x)\right) \end{aligned} \tag{15}$$

A neural operator first encodes the input function $a \in \mathcal{A}$, into one with a larger codomain $h_0 = P(a)$. Then, $h_0$ is passed through $D$ layers which each contains a local operator $W$, an integral kernel $\kappa_\phi$ where the parameter $\phi$ enters and a bias function $b$. Afterwards, $h_{\text{D}}$ is decoded into the output function $u \in \mathcal{U}$ by $u = Q(h_{\text{D}})$. By learning a mapping between function spaces, rather than vector spaces, Neural Operators are *discretization agnostic*, meaning that a model with fixed $\phi$ satisfies the following properties: (1) It can be applied to any discretizations of the input function, (2) it can be evaluated at any point of the output domain and (3) it provably converges to the continuum operator as the number of mesh points is refined (N. Kovachki *et al.* 2023). Similarly to neural networks that compose linear transformations with nonlinear activation functions, neural operators compose linear integral operators with activation functions to achieve universal approximation of nonlinear operators.

FNO is a particular type of Neural Operator that evaluates the kernel integration in the Fourier domain for increased speed (Z. Li *et al.* 2021), that is

$$\begin{aligned} h_{i+1} &= \sigma\left(Wh_i + \mathcal{F}^{-1}\left(\mathcal{F}(\kappa_\phi)\cdot\mathcal{F}(h_i)\right) + b\right) \\ &= \sigma\left(Wh_i + \mathcal{F}^{-1}\left(R_\phi\cdot\mathcal{F}(h_i)\right) + b\right) \end{aligned}, \tag{16}$$

where $W_\phi$ is a linear transformation on $h_i$ and $\mathcal{F}$ and $\mathcal{F}^{-1}$ are Fourier and inverse Fourier transforms, respectively. The encoding $P$, and decoding $Q$, are performed through point-wise multilayer perceptrons (MLPs). Each layer is now defined as a Fourier block (FB). Inside the Fourier block, the representation $h_i$ with large co-domain is convolved with the integral kernel $\kappa_\phi$ parametrized in Fourier spaces, $R_\phi$. This kernel structure is capable of capturing the global information by the nature of convolution, which is also an essential feature that leads to the success of CNNs (Y. LeCun *et al.* 1998). But unlike the CNN, as $R_\phi$ is defined in Fourier space, it is actually a mesh-free integral kernel that is independent of the way the input function is discretized in domain $\Omega$, making its performance substantially better than point-wise mappings (Z. Li *et al.* 2021). However, FNO uses fast Fourier transforms (FFTs) to evaluate Fourier transforms, which restricts the functions to be on regular grids with the same discretization for output $u$ and input $a$; this is not an issue for this paper, and it is important to note that other neural operator architectures are not restricted to regular grids (Z. Li *et al.* 2023; B. Alkin

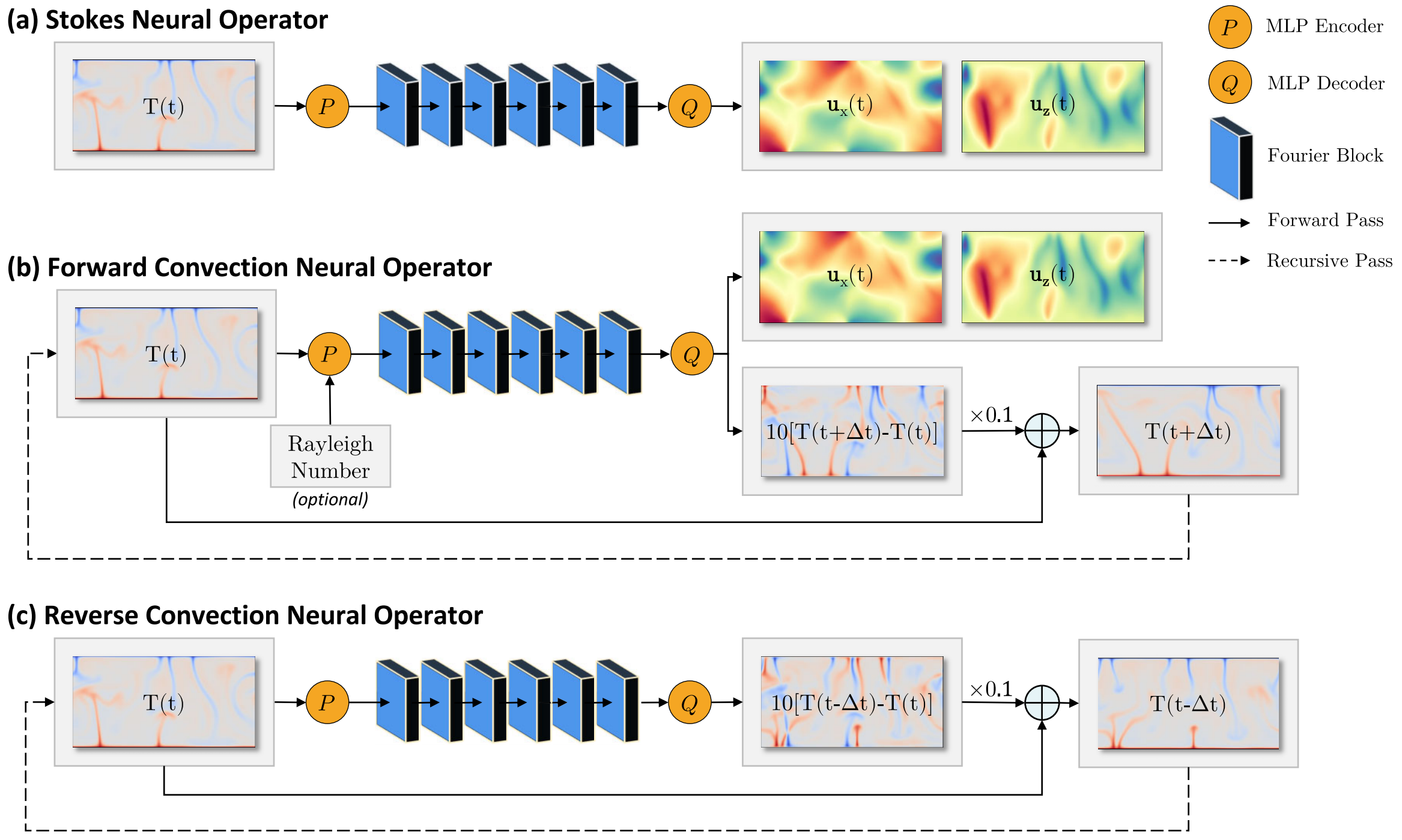

**Figure 2.** The architecture of three neural operators described in this study. Detailed MLP and FB parameters are listed in Table A1.

*et al.* 2024; Y. Shi *et al.* 2025). Since the learning is accomplished in a Fourier space, the resolutions of $g$ and $f$ can be arbitrarily changed during training or evaluation, as long as the grid remains regular. For instance, the neural operator can be trained by loss evaluated on a low resolution mesh, but can be further evaluated on data with higher resolution. This approach is known as super-resolution evaluation (N. Kovachki *et al.* 2023), that both save the training cost and generalize the application scenarios of a trained neural operator, especially for mappings whose underlying physics is scale invariant (Z. Li *et al.* 2021). This feature will also be demonstrated here.

The major architectural hyperparameters that control the complexity and size of the FNO include the number of FB (model depth), the width of FB (model width) and the maximum number of Fourier modes, as FNO does not have to process information on all frequencies in the latent space but can instead make a low-pass cut-off at this mode. To reduce the model size and save on training costs, we adopted an FNO variant, the tensorized Fourier neural operator (TFNO), in which the number of trainable parameters can be greatly reduced by a global factorization on the tensorized parameters (J. Kossaifi *et al.* 2024).

Our first example application of neural operators is as an approximation of the Stokes operator, $\boldsymbol{S}$, which is defined by the Stokes equations (eqs 4 and 5). This is an example of learning an operator explicitly formed by a PDE. In addition, neural operators can also be trained to model hidden relationships between function spaces that lack explicit analytical forms or are intractable to solve numerically. Such procedures are commonly referred to as operator discovery. Some of those mappings could play a role in important connections found in mantle dynamics. A straight-forward example is the forward convection operator $\boldsymbol{F}^{+n\Delta t}$. As described above, the numerical evaluation of this operator requires recursive solutions of the Stokes equation and advection–diffusion equation, while a PDE system that directly maps the initial thermal state to the final one separated by a long time interval does not exist in closed form. The total number of recursive steps depends on the CFL condition, which can be large when modelling full mantle convection problems. The neural operator $\boldsymbol{F}_\phi^{+n\Delta t}$ approximates the mapping between two thermal states over an interval much larger than CFL time steps, which significantly shortens the length of the computational chain and reduces the integration steps required. Such a transformation reduces the time complexity of convection time-integration from $O(N^{3/2})$ with numerical methods to sub-$O(N\log(N))$, effectively dominated by a constant term determined by the size of neural operator, within the considered range of $N$ in this study (see Supplementary Material). It creates cost savings for both forward modelling and the gradient calculation for inverse problems. Another example of model discovery with neural operators is approximating the ill-posed temporal reverse convection operator $\boldsymbol{F}_\phi^{-n\Delta t}$ that maps from the current thermal state to a previous one.

In this study, we will demonstrate how three categories of neural operators (Fig. 2) can be applied to the computation of mantle dynamics: (1) Solving velocity from thermal fields using the Stokes neural operator $\boldsymbol{S}_\phi$, (2) solving the thermal convection problem using $\boldsymbol{F}_\phi^{+n\Delta t}$ and (3) application of neural operators to thermal state reconstruction, including a fast and direct method using $\boldsymbol{F}_\phi^{-n\Delta t}$, and a robust inverse method using $\boldsymbol{F}_\phi^{+n\Delta t}$.

### 2.3.1 *Stokes neural operator*

The Stokes neural operator, $\boldsymbol{S}_\phi$, solves the Stokes equation for velocity and pressure from buoyancy and is the most computationally demanding part of the traditional solution of mantle convection. It takes a temperature field as the source term (input), and outputs velocity and pressure as a response (Fig. 2a). Assuming a constant viscosity, the Stokes operator is linear and its neural approximation is trained with an arbitrary, fixed Rayleigh number and evaluated on systems with other Rayleigh numbers by rescaling outputs.

Similar to other neural architectures, neural operators can be trained using both data-driven and physics-informed approaches. Data-driven methods have a simple form of loss evaluated directly from the deviation between prediction and ground-truth, which can be acquired from forward numerical models. A drawback of this approach emerges when there is insufficient training data or the high-resolution training data is costly to acquire; in such cases, using the known physical laws to formulate the training loss can greatly reduce the demand on the data as well as the pre-training cost (M. Raissi *et al.* 2019; Z. Li *et al.* 2024). For mappings that can be expressed explicitly in terms of PDEs, the physics-informed approach can be conducted by a loss based on the PDE structures. We will demonstrate this by training $\boldsymbol{S}_\phi$ with a purely physics-informed approach. The loss function takes in the physical constraints as

$$L_\mathrm{C} = \|\nabla \cdot \boldsymbol{u}\|_\mathrm{F}, \tag{17}$$

$$L_\mathrm{Mx} = \| \left(-\beta_\mathrm{p}\nabla p + \nabla^2\boldsymbol{u}\right) \cdot \boldsymbol{e}_x\|_F, \tag{18}$$

$$L_\mathrm{Mz} = \| \left(-\beta_\mathrm{p}\nabla p + \nabla^2\boldsymbol{u}\right) \cdot \boldsymbol{e}_z + \beta_\mathrm{u}\mathrm{Ra}(T - T_\mathrm{c})\|_\mathrm{F}, \tag{19}$$

$$L_\mathrm{BP} = \left\| \boldsymbol{u}|_{x=0} - \boldsymbol{u}|_{x=L_x} \right\|_\mathrm{F} + \left\| p|_{x=0} - p|_{x=L_x} \right\|_\mathrm{F}, \tag{20}$$

$$L_\mathrm{BF} = \left\| \boldsymbol{u}_z|_{z=0} \right\|_\mathrm{F} + \left\| \boldsymbol{u}_z|_{z=D} \right\|_\mathrm{F} + \left\| \left.\frac{\partial \boldsymbol{u}_x}{\partial z}\right|_{z=0} \right\|_\mathrm{F} + \left\| \left.\frac{\partial \boldsymbol{u}_x}{\partial z}\right|_{z=D} \right\|_\mathrm{F}, \tag{21}$$

$$L_\mathrm{G} = \left\| \boldsymbol{u}_x|_{x=z=0} \right\|_\mathrm{F} + \left\| p|_{x=z=0} \right\|_\mathrm{F}, \tag{22}$$

$$L_\mathrm{S} = \beta_\mathrm{C}L_\mathrm{C} + \beta_\mathrm{Mx}L_\mathrm{Mx} + \beta_\mathrm{Mz}L_\mathrm{Mz} + \beta_\mathrm{BP}L_\mathrm{BP} + \beta_\mathrm{BF}L_\mathrm{BF} + \beta_\mathrm{G}L_\mathrm{G}, \tag{23}$$

where $L_\mathrm{C}$, $L_\mathrm{Mx}$, $L_\mathrm{Mz}$, $L_\mathrm{BP}$, $L_\mathrm{BF}$, $L_\mathrm{G}$ and $L_\mathrm{S}$ are continuity, horizontal momentum, vertical momentum, periodic boundary, free-slip boundary, gauge (null-space constraints fixing the additive constant of pressure and eliminating the net horizontal velocity mode) and total Stokes PDE losses, respectively, and $\|\cdot\|_\mathrm{F}$ is the Frobenius norm. The gradient optimization forces the neural operator to generate $\boldsymbol{u}$ and $p$ that satisfy conservation and boundary conditions. As our quantities of interest are the velocity fields, the buoyancy term in eq. (19) is written with a constant temperature $T_\mathrm{c} = 0.5$ subtracted. This only introduces a constant vertical pressure gradient, and was found in practice to improve the training of physics-informed neural operator (PINO). The discretizations of $L_\mathrm{C}$, $L_\mathrm{Mx}$ and $L_\mathrm{Mz}$ are shown in Fig. B1 and eqs (B1)–(B3). Here we present the PINO losses in their continuum form. Their discrete evaluation is based on staggered finite-difference stencils given in Fig. B1 and eqs (B1)–(B3), which are straightforward to implement and efficient on tensor-based GPU workflows (Z. Li *et al.* 2024) (see Supplementary Material). For problems posed on unstructured meshes, the PDE residual can alternatively be formulated and evaluated in a variational (weak) form (E. Kharazmi *et al.* 2019). The weighting coefficients, $\beta_\mathrm{C}$, $\beta_\mathrm{Mx}$, $\beta_\mathrm{Mz}$, $\beta_\mathrm{BP}$, $\beta_\mathrm{BF}$ and $\beta_\mathrm{G}$ are applied on each loss term to balance the weights between conservation and boundary conditions. In eqs (17)–(23), $p$ and $\boldsymbol{u}$ denote the normalized raw output channels of $\boldsymbol{S}_\phi$, whose scaling, $\beta_\mathrm{u}$ and $\beta_\mathrm{p}$, are chosen to make the optimization better conditioned (values listed in Table B1). The corresponding true dimensionless pressure and velocities can be recovered from the normalized raw output channels of $\boldsymbol{S}_\phi$ as $p_\mathrm{true} = (\beta_\mathrm{p}/\beta_\mathrm{u})p_\mathrm{raw}$ and $\boldsymbol{u}_\mathrm{true} = \boldsymbol{u}_\mathrm{raw}/\beta_\mathrm{u}$. With this rescaling convention, the zero-PDE-residual training condition remains unchanged. Accordingly, all $p$ and $\boldsymbol{u}$ appearing in the detailed PDE loss stencil equations (eqs B1–B3) also denote the normalized raw output fields.

By adding a data loss term to eq. (23), we can construct a hybrid training approach that uses both existing data and PDEs for training $\boldsymbol{S}_\phi$. However, we show model performance under the end-member condition where no training data is available; thus, $\boldsymbol{S}_\phi$ here is a purely physics-informed model without using any synthesized training data. As the loss function of $\boldsymbol{S}_\phi$ is data-free, the only pre-training computation needed are the random input thermal field $T$. Here, we use the random field with an exponential covariance kernel

$$K(x, x') = \exp\left(-\frac{|x - x'|}{2l}\right) \tag{24}$$

whose variogram has a broad effective range of $|x - x'|$ (a long tail in its variogram function) compared with other common covariance models used in geostatistics, such as Gaussian or spherical kernels (R. Webster & M.A. Oliver 2007; S. Müller *et al.* 2022). We utilize this property to incorporate structures of different wavelengths in our input so as to stimulate the operator learning effectively on all modes.

*2.3.2 Forward and reverse convection neural operators*

The convection neural operators $\boldsymbol{F}_{\phi}^{\pm n\Delta t}$ solve the forward $(+n\Delta t)$ and reverse $(-n\Delta t)$ convection problems within domain $\Omega$ with a fixed Rayleigh number described by eqs (4)–(6). $\boldsymbol{F}_{\phi}^{+n\Delta t}$ approximates the mapping from a thermal state $T(\mathbf{x}, t)$ to both the Stokes velocity solution $\boldsymbol{u}(\boldsymbol{x}, t)$ and to another state that is $n\Delta t$ ahead in time, that is, $T(\mathbf{x}, t + n\Delta t)$ (Fig. 2b), while $\boldsymbol{F}_{\phi}^{-n\Delta t}$ approximates the mapping from a thermal state $T(\boldsymbol{x}, t)$ to a previous state at $n\Delta t$ ago (Fig. 2c).

For the convection neural operators, either the forward or the reverse, their outputs and inputs are not explicitly connected by differential relations, or PDEs, rather a data-driven approach is used for training. An input–output training data pair is sampled from the thermal convection process (eqs 4 and 5) computed with `Underworld`. Due to the dissipative characteristic of $\boldsymbol{F}^{+n\Delta t}$, the simplest way to create the training data set spanning the input function space $\mathcal{T}$ is to integrate the thermal states from some initial random fields. Here, the initial states of the training data set are random thermal fields satisfying Dirichlet boundary conditions with a Gaussian covariance kernel:

$$K(x, x') = \exp\left(-\frac{(x - x')^2}{2l^2}\right). \qquad (25)$$

Unlike the exponential covariance model (eq. 24), the variogram of a Gaussian covariance model increases most rapidly near $|x - x'| = l$ which reflects the characteristic length scale within the thermal structure over which spatial correlation decays. Since the temperature at the upper boundary is fixed, such a characteristic correlation length scale will form an initial thermal boundary layer with thickness $l$. The thermal profile shows that this Gaussian boundary thermal gradient is similar to the error-function thermal gradient produced by conduction. We set $l$ slightly larger than the boundary layer thickness at the steady state, in order to ensure that the overturn phase occurs quickly, incorporating abundant random movement, while avoiding the possible numerical instability brought on by large deviation from the distribution of $\mathcal{T}$.

From each Gaussian initial thermal field, a temporal sequence of convection data can be generated by forward integration, and a large thermal convection data set can thus be created from multiple random initial states. By sampling data pairs with a specific time interval, indicated by the training data integration time scope shown in Fig. 1, we constructed the training data set for forward convection neural operators (Fig. S4, Supplementary Material). Reverse convection neural operators can be trained with exactly the same strategy by exchanging the previously sampled input–output pairs.

In addition to accelerating the forward computation, neural operators are also widely used in inverse problems (e.g. Z. Li *et al.* 2021; C. Zou *et al.* 2025). When the forward calculation is performed as a sequential forward pass through the neural operator, a traceable computational graph is found with gradients between nodal values, thus the accumulated derivatives can be calculated using the chain rule, either from inside to outside (forward accumulation) or from outside to inside (reverse accumulation/backpropagation). This technique is known as automatic-differentiation and circumvents the need of solving adjoint equations (A. Griewank & A. Walther 2008), and can be easily performed in a highly parallel way on GPUs. In fact the adjoint-state method is mathematically equivalent to automatic-differentiation (W. Zhu *et al.* 2021). Hence, the gradient is now calculated using backpropagation and the derived gradient is more accurate compared to solving an adjoint equation, because it does not incorporate any numerical errors and can reach machine precision if the forward model is assumed as an accurate approximation to the ground-truth operator.

## 3 RESULTS

### 3.1 Physics-informed Stokes neural operator $\mathbf{S}_{\phi}$

The Stokes equations (eqs 4 and 5) describe the mapping from the buoyancy field and viscosity structure to velocity and pressure, denoted as the Stokes operator $\boldsymbol{S}$. In its neural operator form $\boldsymbol{S}_{\phi}$ is trained with a purely physics-informed approach with loss function, $L_{\mathrm{S}}$ (eq. 23). In this instance, we test the neural operator's capability of a super-resolution prediction (Z. Li *et al.* 2021): $\boldsymbol{S}_{\phi}$ is trained with random thermal inputs discretized on lower-resolution meshes ($129 \times 129$) and later evaluated on a finer one ($257 \times 257$). First, we apply $\boldsymbol{S}_{\phi}$ to the temperature fields sampled from time window 1 (Fig. 1). For this forward convection example at $\mathrm{Ra} = 10^7$, we found that it successfully resolves the flow even with high-frequency plume structure and achieves a $L_2$ relative error of $\sim$5 per cent compared to the velocity components solved by `Underworld`. Then, we also tested the accuracy of $\boldsymbol{S}_{\phi}$ by using three convecting thermal structures produced with $\mathrm{Ra} = 10^5, 10^6, 10^7$, and different resolutions (65, 129, 257) as inputs (Fig. S1, Supplementary Material) and they performed equally well with their $L_2$ relative errors all around 3–5 per cent (Table 1). The point-wise errors in general align with the amplitudes of velocity, showing some long wavelength features where the largest error occurs at the edges of convection cells, and diminishes within the thermally homogeneous regions. For $\boldsymbol{u}_x$, it occurs within the upper and lower thermal boundaries where the horizontal velocities reach the maximum; for $\boldsymbol{u}_z$, it occurs near the downwellings and upwellings. In general, the convection velocity patterns predicted by $\boldsymbol{S}_{\phi}$ are less vigorous with lower amplitudes. However, $\boldsymbol{S}_{\phi}$ seems to be capable of predicting accurate velocities within the thermal cores of the upwellings and downwellings. Often it is difficult to analytically explain the origin behind prediction errors of a neural network, however, the features here seem to indicate that the Stokes neural operator resolves the local response of a thermal anomaly, but lacks the ability to accurately reproduce the spatial response in neighbour regions, compared with neural operators trained with data-driven methods, as detailed below. As there is usually a trade-off between computational cost and solver accuracy, the moderate 5 per cent prediction error could still make the physics informed Stokes neural operator usable for some forward, mantle dynamic computations, as considered in the Discussion.

**Table 1.** Relative prediction errors of Stokes neural operator with convection patterns under different Rayleigh numbers. For $\boldsymbol{u}_x$, the spatial mean is removed prior to computing the relative $L_2$ error due to the null mode associated with periodic lateral boundary conditions. The same treatment is applied in Table 2.

| Neural operator | Ra | $L_2$ Relative error | |
|---|---|---|---|
| | | $\boldsymbol{u}_x$ | $\boldsymbol{u}_z$ |
| $\boldsymbol{S}_\phi$ | $10^5$ | 5.07 per cent | 4.35 per cent |
| | $10^6$ | 3.76 per cent | 3.32 per cent |
| | $10^7$ | 5.30 per cent | 4.81 per cent |

**Table 2.** The symbols of forward (+) and reverse (−) convection neural operators, their integration times under different Rayleigh numbers and one-step prediction error levels of forward convection neural operators.

| Neural operator | Ra | $n\Delta t$ | Forward model single step $L_2$ relative error | | |
|---|---|---|---|---|---|
| | | | $T$ | $\boldsymbol{u}_x$ | $\boldsymbol{u}_z$ |
| $\boldsymbol{F}_{\phi_5}^{\pm 1}$ | $10^5$ | $2.5 \times 10^{-3}$ | 3.82 per cent | 0.06 per cent | 0.06 per cent |
| $\boldsymbol{F}_{\phi_6}^{\pm 1}$ | $10^6$ | $2.5 \times 10^{-4}$ | 0.67 per cent | 0.04 per cent | 0.04 per cent |
| $\boldsymbol{F}_{\phi_7}^{\pm 1}$ | $10^7$ | $2.5 \times 10^{-5}$ | 0.13 per cent | 0.07 per cent | 0.10 per cent |
| $\boldsymbol{F}_{\phi_5}^{\pm 2}$ | $10^5$ | $5 \times 10^{-3}$ | 5.25 per cent | 0.06 per cent | 0.05 per cent |
| $\boldsymbol{F}_{\phi_6}^{\pm 2}$ | $10^6$ | $5 \times 10^{-4}$ | 1.38 per cent | 0.05 per cent | 0.07 per cent |
| $\boldsymbol{F}_{\phi_7}^{\pm 2}$ | $10^7$ | $5 \times 10^{-5}$ | 0.29 per cent | 0.13 per cent | 0.11 per cent |
| $\boldsymbol{F}_{\phi_5}^{\pm 4}$ | $10^5$ | $1 \times 10^{-2}$ | 4.58 per cent | 0.10 per cent | 0.09 per cent |
| $\boldsymbol{F}_{\phi_6}^{\pm 4}$ | $10^6$ | $1 \times 10^{-3}$ | 2.88 per cent | 0.06 per cent | 0.06 per cent |
| $\boldsymbol{F}_{\phi_7}^{\pm 4}$ | $10^7$ | $1 \times 10^{-4}$ | 0.87 per cent | 0.16 per cent | 0.17 per cent |

As for the computational costs of formulating such a surrogate model, $\boldsymbol{S}_\phi$ is a purely physics-informed neural operator that does not require any pre-training computational resource on solving the Stokes equations with traditional numerical methods to generate the training data set. Rather, with random source term inputs, the training processes start at no extra costs by evaluating several PDE losses from the output end, and the evaluation of PDE losses is much simpler than deriving the solution, especially when the system becomes nonlinear and an iterative method is required to derive the solution. In other words, we achieved to make the entire workflow of training the Stokes neural operator and solving the forward problem in a 'CPU-free' manner. Besides, we utilized the super resolution feature of neural operators, that we trained $\boldsymbol{S}_\phi$ on a relatively coarse mesh, which further reduces the GPU expense. Hence, we suggest that for a one-step prediction model of which the underlying physics is already prescribed explicitly, the physics-informed approach could be the optimized strategy because of its efficiency.

### 3.2 Data-driven forward convection neural operators $\boldsymbol{F}_\phi^{+n\Delta t}$

We now demonstrate solving the forward convection problem using neural operators $\boldsymbol{F}_\phi^{+n\Delta t}$ with three different integration times (symbols listed in Table 2) over time window 1 in the computed convection sequence (Fig. 1), spanning about 14 transit times ($t_{tr}$), with a Rayleigh number of $10^7$. This test is carried out to evaluate the neural operators' accuracy and stability in modelling long-term convection dynamics. Compared with the CFL limit in traditional solvers, the integration time steps of forward convection neural operators are about hundred times larger. For $\boldsymbol{F}_{\phi_7}^{\pm 4}$ (the one with the largest time step), this contrast can reach ∼300 times (Tables 2 and C1).

The operator $\boldsymbol{F}_\phi^{+n\Delta t}$ can be called recursively to compute time-dependent convection by integrating a thermal field forward to a next time step. It also has another output channel that predicts the current step velocity components. Comparisons between the output snapshots of thermal and velocity fields from forward computations using $\boldsymbol{F}_{\phi_7}^{+n\Delta t}$ and `Underworld` have been made (Fig. 3). In addition, diagnostic variables mid-depth Nusselt number, Nu, and the maximum horizontal velocity, $\boldsymbol{u}_{x0}$, are also tracked within the time window for comparison.

All three forward convection neural operators with different integration time steps produce results that track almost equally well with the sequence computed by `Underworld`. The differences between the two methods remain small through the first transit time in terms of the internal fields and Nu and $\boldsymbol{u}_{x0}$. The relative error of neural operators' one-step prediction can be as low as 0.1 per cent (Table 2). After 3-4 transit times, the states computed by neural operators gradually diverge with those from `Underworld`, which is manifested in Nu and $\boldsymbol{u}_{x0}$ trajectories. This phenomenon is expected and almost inevitable as the Ra $= 10^7$ system is highly nonlinear, such that a small mismatch in the initial stages of the computation cause increasing deviations during the forward integration. Although the detailed convection patterns might deviate, both methods bring the convection process into a relatively steady state with two unit aspect ratio cells after 8 transit times, which indicates that the neural operators successfully learn the dissipative dynamics, reproducing an attractor near a steady state with two near-stationary convection cells during long-term computation, even if the training data pairs only contain the non-steady convection patterns (Fig. S4, Supplementary Material).

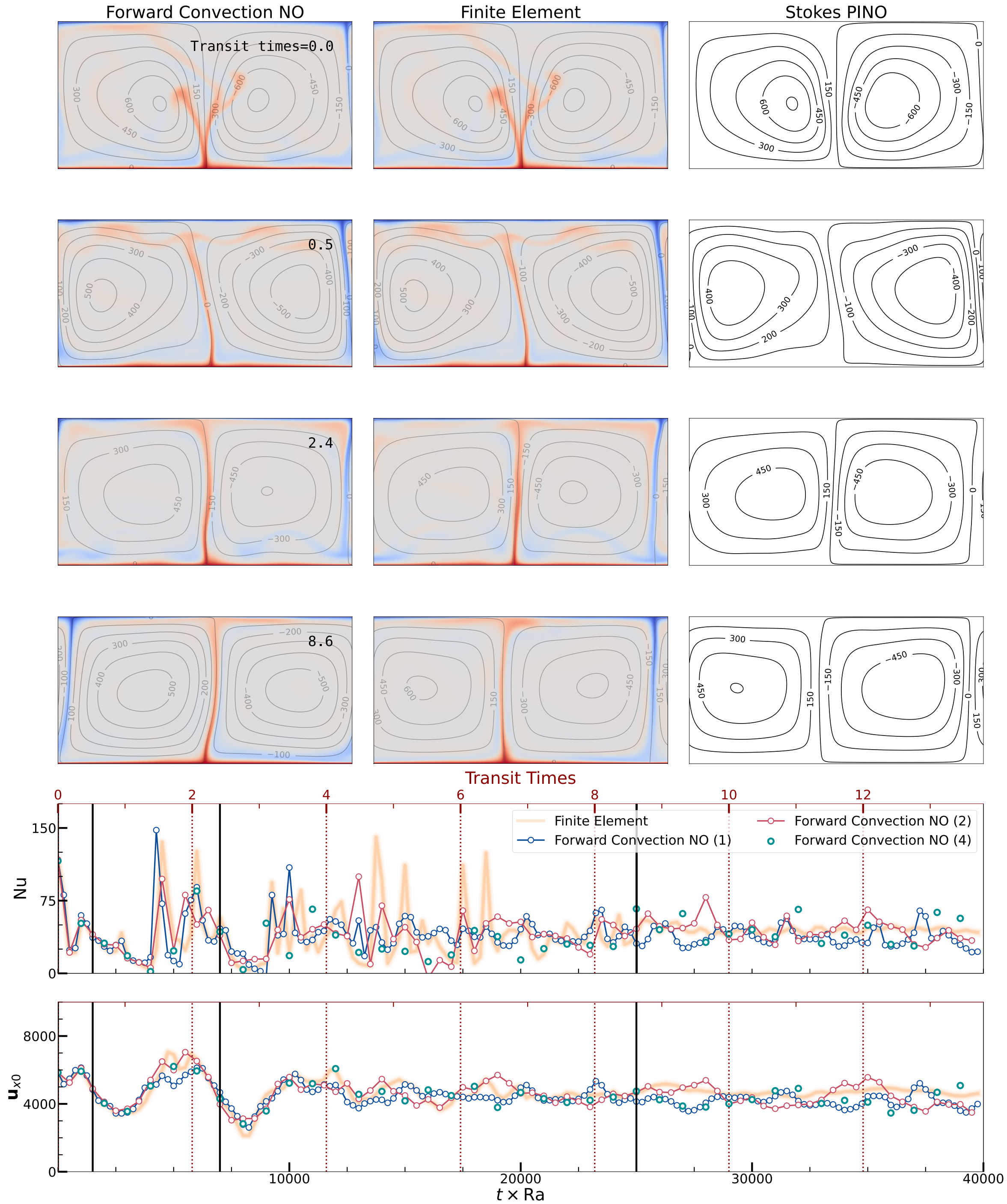

**Figure 3.** Comparisons between forward computations using $\boldsymbol{F}_{\phi_7}^{+n\Delta t}$ and `Underworld`, Ra $= 10^7$. Rows 1 to 4 show the forward evolution snapshots of the systems. Column 1: temperature snapshots and velocity streamlines computed by $\boldsymbol{F}_{\phi_7}^{+1}$, system integrated by $\boldsymbol{F}_{\phi_7}^{+1}$; Column 2: temperature and velocity computed by `Underworld`, system integrated by `Underworld`; Column 3: velocity computed by $\mathcal{S}_\phi$ based on thermal fields integrated by `Underworld` as inputs. Rows 5 and 6: Nu and $\boldsymbol{u}_{x0}$ tracked in systems integrated with different methods. Black short bars indicate the instants where the snapshots within rows 1 to 4 are chosen. For consistency of comparison, the horizontal means of all $\boldsymbol{u}_x$ are adjusted to zero. The original Stokes PINO employs a pointwise gauge condition ($\boldsymbol{u}_x|_{x=z=0} = 0$), whereas the other two models follow a zero-mean gauge convention under periodic boundary conditions. The same treatment is applied in Figs S5 and S6 (Supplementary Material).

We also investigated how the neural operators' integration time step size influences the prediction accuracy. Although a clear trend of temperature prediction error with time step size is observed for $\boldsymbol{F}_{\phi_7}^{+n\Delta t}$, we found that a smaller time step does not guarantee a more precise prediction when integrating a thermal field over a specific, long duration of time. The number of recursive steps is inversely proportional to the integration time, which results in a trade-off. By comparing against temperature predictions from `Underworld`, we find that for accuracy in long-term forward integration, the three forward operators have a similar performance. On the other hand, as the velocity predictions are solutions of an instantaneous Stokes problem and thus independent of the forward integration time step size, its prediction error remains less than 0.1 per cent (Table 2). In addition to the mean one-step errors reported in Table 2, Fig. S2 (Supplementary Material) summarizes the distribution of one-step prediction errors over time window 1. Across all three operators, the error histograms are sharply peaked, indicating that the mean values in Table 2 are representative of typical behaviour rather than being driven by a small number of anomalous snapshots. All three channels exhibit some extent of upper-tail events; however, the tails are notably heavier for the velocity channels than for temperature. This is favourable for our autoregressive rollouts, since only the predicted temperature is fed back to advance the state, whereas the Stokes velocities are evaluated diagnostically at each step and are not used as inputs for subsequent predictions.

Nevertheless, circumventing the CFL constraint while using GPU computation, all three surrogate models accelerate the forward integration substantially with speedups proportional to the neural operators' time step size. For this problem, $\boldsymbol{F}_{\phi_7}^{+1}$, $\boldsymbol{F}_{\phi_7}^{+2}$, $\boldsymbol{F}_{\phi_7}^{+4}$ achieve speedups of around 5000×, 10000× and 20000×, respectively, with a mesh of 257 × 257 (computational costs and comparisons listed in Table S3 in Supplementary Material). The speedup factor could even scale up as the system size becomes larger (see Discussion and Supplementary Material for more details).

The proposed forward convection neural operator architecture shows good performance under different Rayleigh numbers. We first trained forward convection neural operators using the same architecture as $\boldsymbol{F}_{\phi_7}^{+n\Delta t}$, with training data with Ra = $10^6$ and $10^5$ (Figs S5 and S6, Supplementary Material). In lower Rayleigh number cases, the thermal convection is less unsteady and the underlying physics is easier for the neural operators to approximate in terms of the long-term prediction errors. We find that for all the forward convection neural operators with different integration time steps tested (Table 2), predictions match and track the computation from `Underworld` more accurately as compared with the Ra = $10^7$ case, and hold for tens of transit times. The performance of each forward convection neural operator is evaluated by inputting all of the ground-truth thermal states from the complete convection sequences (such as the one shown in Fig. 1) under three Rayleigh numbers to the corresponding neural operators and then calculated the relative $L_2$ error of each variable against the `Underworld` results. By averaging over the entire time window, we derived one-step prediction errors of all the output quantities (Table 2). They all remain small, implying that the forward convection neural operators are all well trained. Their accuracy in modelling the thermal convection through several transit times indicates their capability of becoming suitable surrogate forward models for inverse problems. We further assessed the fidelity of our data-driven operator discovery workflow by benchmarking low-Rayleigh-number cases against the steady isoviscous convection results of B. Blankenbach *et al.* (1989) at Ra = $10^5$ and $10^6$ (Table C2). Since the benchmark is defined in a unit-aspect-ratio box with free-slip boundaries, we generated training data and trained neural operators in the same geometry and boundary conditions as the benchmark. We report the surface Nusselt number, root-mean-square (RMS) velocity, and corner heat fluxes, with relative errors of approximately 3, 1 and 5 per cent, respectively, for both Rayleigh numbers. With the exception of RMS velocity, the errors increase with Rayleigh number. This trend is consistent with the single-step prediction results in Table 2: That as Ra increases, steady convection develops progressively finer thermal structures, which are more challenging for the surrogate to resolve and can lead to larger errors in thermodynamic diagnostics (e.g. temperature and heat flux) than in kinematic quantities, as the Stokes outputs are generally concentrated in long wavelength. This is also consistent with prior observations that Fourier-based architectures can have difficulty in the presence of sharp gradients or high-frequency content (Z. Li *et al.* 2021).

So far, the forward convection neural operators have been trained and evaluated at a fixed Rayleigh number, matching the parameter value used to generate the training data. To demonstrate that the surrogate can incorporate scalar PDE parameters and learn their impact on the dynamics, we additionally trained a parameter-conditioned forward operator, denoted by $\boldsymbol{F}_{\phi_{\mathrm{Ra}}}^{+1}$, which takes continuous Rayleigh number as an extra input channel and performs evaluation on the temperature field and velocity fields as well. The Rayleigh number roughly measures the ratio of buoyancy driving to viscous friction, and controls convective vigour and heat transport efficiency. In classical Rayleigh–Bénard convection, boundary layer theory (BLT) predicts a power-law relation $\mathrm{Nu} \propto \mathrm{Ra}^{1/3}$ over appropriate parameter ranges (D.L. Turcotte & G. Schubert 2002). We therefore test whether $\boldsymbol{F}_{\phi_{\mathrm{Ra}}}^{+1}$ can reproduce the Nu–Ra scaling over $\mathrm{Ra} \in [10^4, 10^8]$ (Fig. 4). This interval includes the training range $\mathrm{Ra} \in [10^5, 10^7]$ and two extrapolation regions outside it. Within the training range, $\boldsymbol{F}_{\phi_{\mathrm{Ra}}}^{+1}$ yields a log–log slope of ≈0.30, close to the expected 1/3 scaling, and achieves error levels comparable to the fixed-Ra operators and the reference numerical solutions. Beyond the training range, the model exhibits partial zero-shot generalization: For $\mathrm{Ra} < 10^5$, the predicted scaling remains close to 1/3 for roughly half an order of magnitude before deviating, whereas for $\mathrm{Ra} \gtrsim 10^7$, the slope departs more notably from 1/3. The larger extrapolation error at high Ra is consistent with increasingly complex convection, which is more challenging to represent with a band-limited Fourier-based architecture. We therefore treat those points as outside the reliable operating range of $\boldsymbol{F}_{\phi_{\mathrm{Ra}}}^{+1}$ and focus our quantitative assessment on the regimes where the scaling behaviour remains physically consistent.

### 3.3 Thermal state reconstruction

The outcomes from four different methods for the thermal state reconstruction (two backward integrations and two inverse methods) were computed and their performances were compared. A convection sequence of time window 3 (Fig. 1) computed with `Underworld` lasting for about 1 transit time is used as the ground-truth data against which the reconstructions are compared (Fig. 5, Column 1). To compute a realistic problem setup, our synthetic observables include the terminal thermal state and horizontal velocities on the upper surface as a function of time (Fig. 5, Column 6). Two backward integrations reverse convection directly from the terminal thermal state without utilizing the surface velocity data. As for the two inverse methods, we first incorporate observations from the terminal thermal state only, which takes the identical information as the direct methods; separate joint inversion is conducted while also using surface velocity observations. The reconstruction target is the initial thermal state (Figs 5 and 6a). When evaluating the robustness of each method, ultimately for geophysical inversion, we consider a second experiment where the observations are polluted with noise (Figs 7 and 6b).

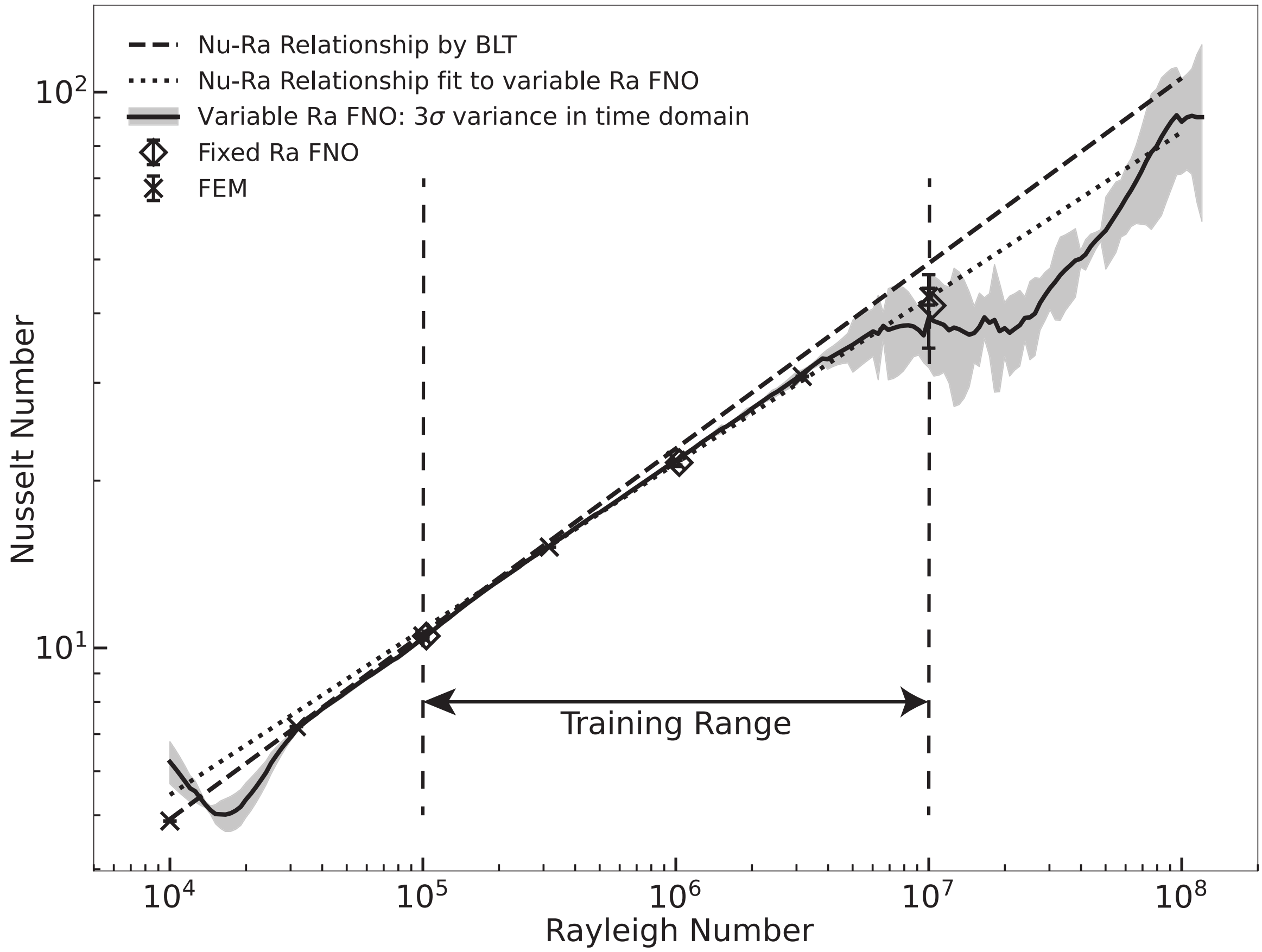

**Figure 4.** Rayleigh number–Nusselt number scaling of the variable Rayleigh number convection operator $\boldsymbol{F}^{+1}_{\phi_{\mathrm{Ra}}}$, compared against: the fixed Rayleigh number convection neural operator $\boldsymbol{F}^{+1}_{\phi_5}$, $\boldsymbol{F}^{+1}_{\phi_6}$, $\boldsymbol{F}^{+1}_{\phi_7}$; boundary layer theory; and results from the finite element method. The training range of Ra is $[10^5, 10^7]$. Beyond this scope, zero-shot extrapolations test the ability of predicting $\boldsymbol{F}^{+1}_{\phi_{\mathrm{Ra}}}$ from Ra $= 10^4$ to $10^8$. The coefficient of BLT Nu–Ra 1/3 power scaling law is $\left(\frac{\lambda}{2D}\right)^{2/3} / \left[2^{1/3}\pi^{2/3}\left(1+\frac{\lambda^4}{16D^4}\right)^{1/3}\right]$, with aspect ratio of convection cells $\lambda/2D = 1$, $\lambda$ the wavelength and $D$ the box depth (D.L. Turcotte & G. Schubert 2002).

The evolution in time window 3 starts from an initial state where there exists one upwelling and one downwelling plume spanning the domain. The system is not yet in a steady convective state (Fig. 1), which can be gleaned from the initial temperature field with several boundary layer instabilities. Forward in time, these thermal instabilities grow at the thermal boundary layers, velocities increase and growing plumes merge into the existing larger ones (Fig. 5, Column 1, dynamics studied in detail in earlier literature, U. Hansen & A. Ebel (1988)). The surface velocity profile notably changes during the process, where the amplitude of convergence and divergence increases as new thermal structures mature; the imprints of those newly initiated structures are clearly recorded by the surface velocity profiles (Fig. 5, Column 6). Moreover, the non-stationarity of the convection is evident in the final complex thermal structures, with plumes bent from the vertical and with distinct changes in plume width as a function of depth. Such complexity in the thermal structure retains information on the earlier time-dependence of the flow which we now attempt to recover.

We begin with the direct method of reversing the direction of buoyancy. Through backwards integration in time with reverse buoyancy and allowing diffusion to operate normally backwards in time (eqs 8, 9 and 11) (Fig. 5, Column 2) a crude recovery is achieved. In this case, the thermal structures are advected back to the thermal boundary layers, rather than recovering the previous structures evident from the signals preserved in the current state. The consequence is that the reconstructed thermal state eventually becomes homogeneous with thick diffusive boundary layers and this method only recovers structures to less than $t = -0.34t_{\mathrm{tr}}$, confirming earlier results (C.P. Conrad & M. Gurnis 2003; A. Ismail-Zadeh *et al.* 2004).

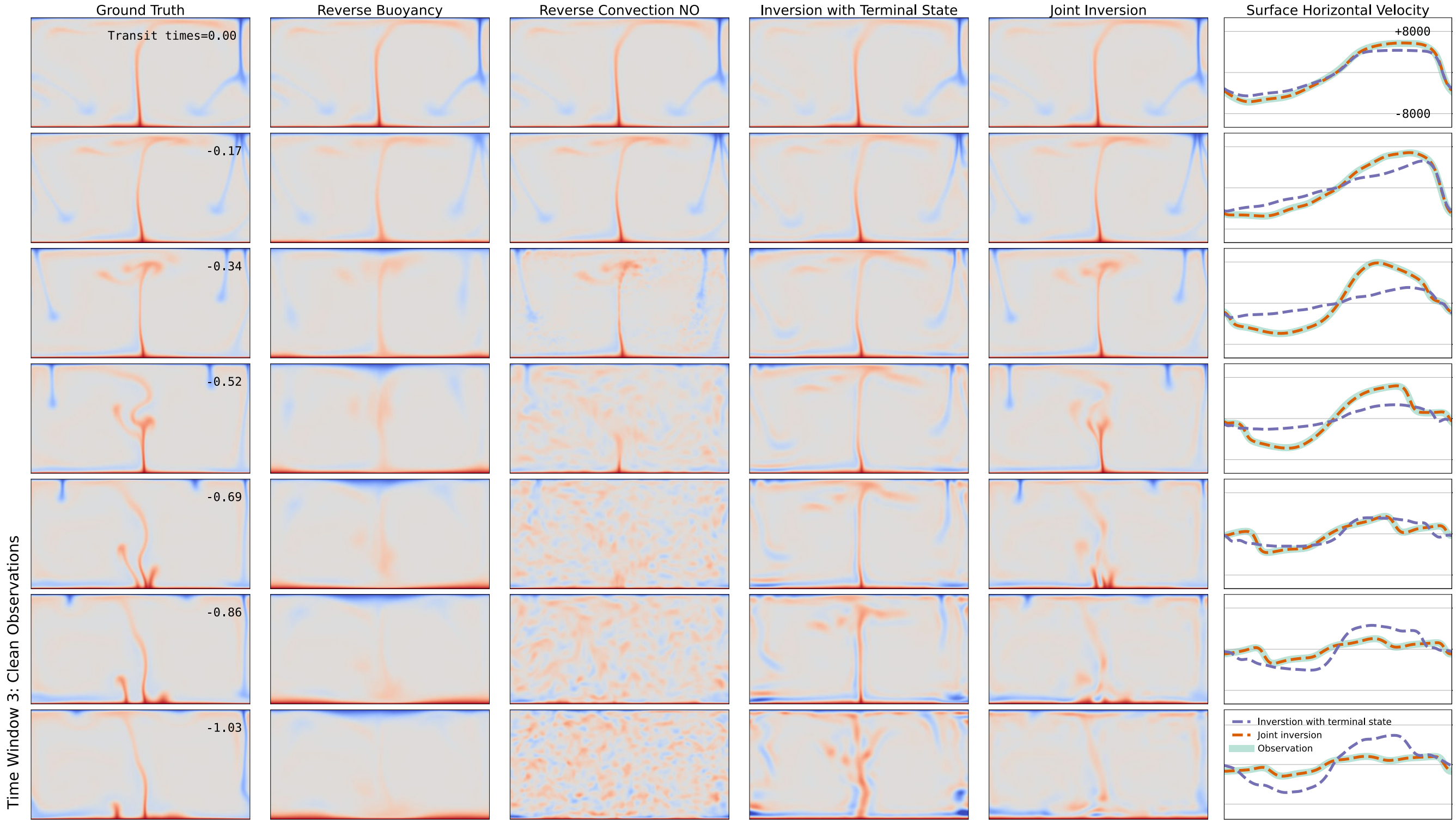

**Figure 5.** Reconstruction performances of different methods over time window 3 with clean observations. Reversal time steps shown by row (from top to bottom backwards in time). Column 1: A ground-truth thermal convection sequence within time window 3 computed by `Underworld` forward in time from bottom to top; Columns 2–5: Reconstruction with: reverse buoyancy; reverse convection operator; inversion with terminal state only; joint inversion; Column 6: At each time step, the surface horizontal velocity profiles compare the clean observations with predictions from the inversion with terminal state only and the joint inversion.

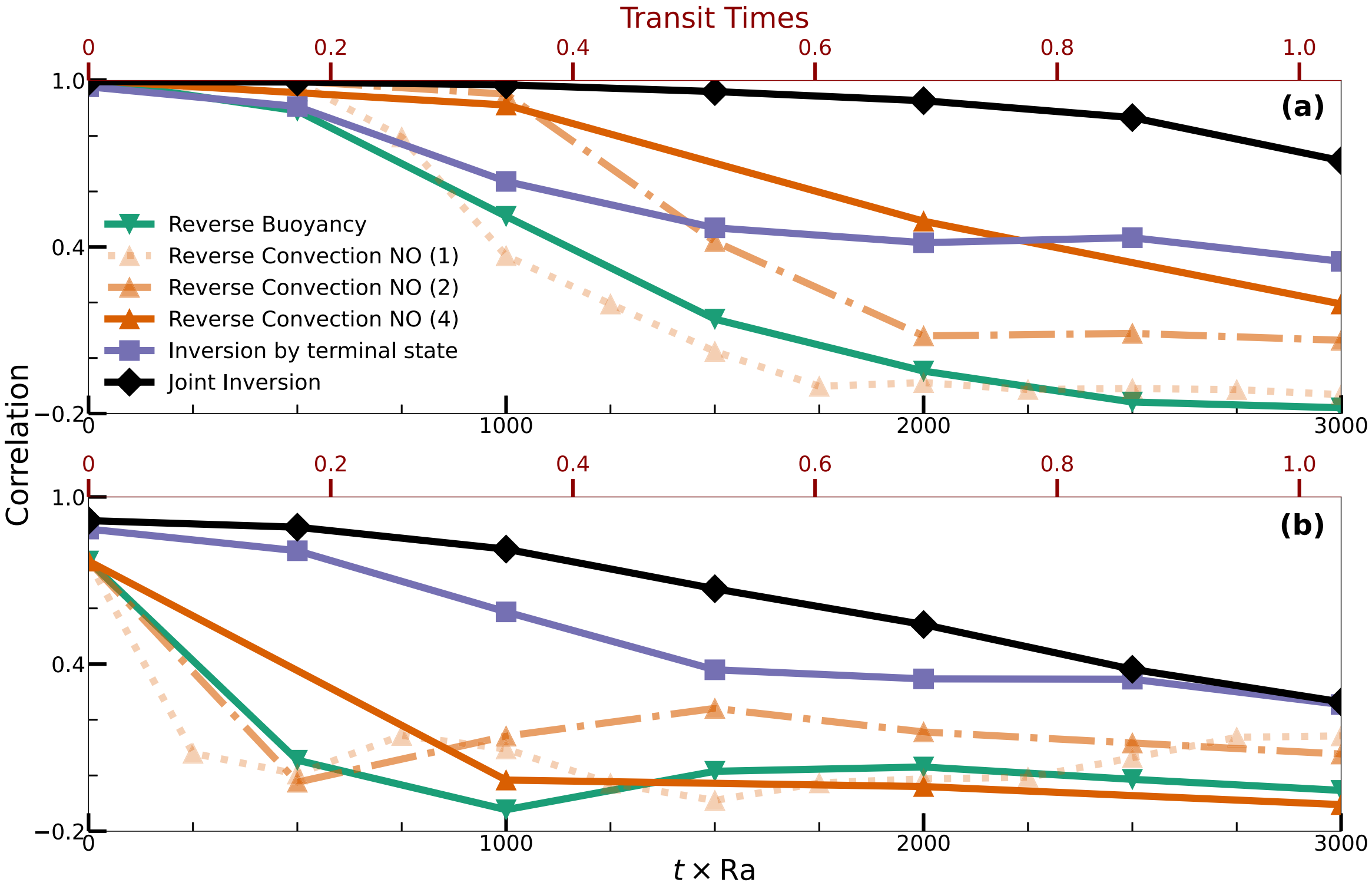

**Figure 6.** Correlation coefficient of reconstructed thermal fields with ground-truth fields versus backward time. Each line represents a different reconstruction method. (a) Reconstruction with synthesized observations (no noise). (b) Reconstruction with synthesized observations polluted with 5 per cent pink noise.

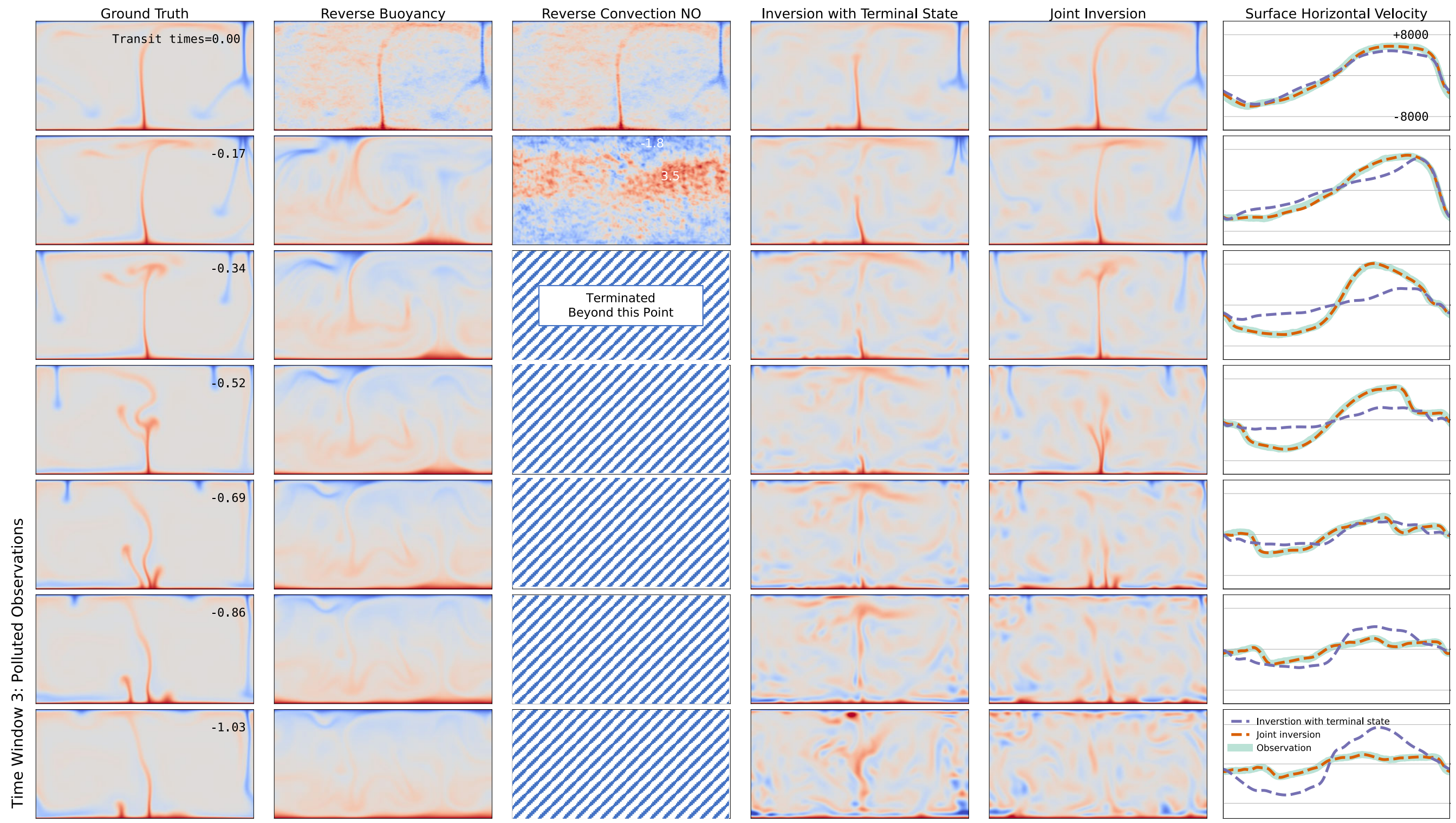

**Figure 7.** Reconstruction performances of different methods over time window 3 with polluted observations. Reversal time steps shown by row (from top to bottom backwards in time). Column 1: A ground-truth thermal convection sequence computed by `Underworld` forward in time from bottom to top; Columns 2–5: Reconstruction with: reverse buoyancy; reverse convection operator; inversion with terminal state only; joint inversion; Column 6: At each time step, the surface horizontal velocity profiles compare the observations polluted with 5 per cent pink noise with predictions from the inversion with terminal state only and the joint inversion.

The second direct method integrates backward using the reverse convection neural operator (Fig. 5, Column 3). Compared with reverse buoyancy integration, the reverse convection neural operator not only recovers the dominant thermal structures with the correct amplitudes, but also recovers structure that subsequently underwent substantial diffusion. At $t = -0.34t_{\mathrm{tr}}$, although we find that both methods reconstruct the merging of the downwelling on the right side of the box, only the reverse convection neural operator clearly recovers the downwelling on the left side, which had dissipated by the current state. By reversing buoyancy only, the vague signals of this mostly vanished structure are substantially smoothed. The fundamental difference between the two methods is that the reverse convection neural operator learns the anti-diffusion operator $-\nabla^2$, while by reversing buoyancy only, diffusion is over estimated during the backward integration. Consequently, the reverse convection neural operator is able to reconstruct further backwards in time. In contrast to the forward operator, we observe that the integration time step size of the reverse operator has a large influence on the accuracy of the recovered temperature over a specific long duration of time; for example, $\boldsymbol{F}_{\phi_7}^{-4}$ (Fig. 5, Column 3) with the largest integration time step outperforms the others with smaller ones (Fig. 6). As $\boldsymbol{F}_{\phi_7}^{-4}$ has such a large step size, the reconstruction does not exactly match the ground truth steps shown, and consequently we integrate forward in time using forward convection neural operators between the check points to display a complete snapshot sequence in Fig. 5.

The inversion to reconstruct the initial state constrained only by the terminal temperature field is explored next (Fig. 5, Column 4). Compared with direct methods, the inversion recovers a convection pattern further back in time with quite realistic cold downwellings and warm plumes in the interior compared with direct methods. However, thermal structures recovered before $t = -0.52t_{\mathrm{tr}}$ appear to be non-physical with strong cold halos characterized by strong gradients around hot plumes and vice-versa, which lead to an incorrect evolutionary path. One possibility is that the optimization becomes trapped in non-physical, local minima by this time.

Finally, we utilized the information from both the terminal thermal state and surface velocity profiles in a joint inversion (Fig. 5, Column 5). We find that this approach outperforms all others by successfully reconstructing almost all thermal structures that once existed within the time window, including the merging of plumes and the onset of instabilities. Particularly, the entire life cycle of the downwelling on the left side of the box is recovered. In addition, the reconstruction resolves the three upwellings that existed prior to their merger. The thermal field recovered at $t = -1.03t_{\mathrm{tr}}$ (over one transit time) by joint inversion is still informative and contains the correct long wavelength structure compared with the ground truth. The initial state recovered by joint inversion leads to an evolutionary sequence that matches both the terminal states and the surface velocity profiles well (Fig. 5, Column 6).

Defined sequentially on the time axis, the chronological data stabilizes the inversion backward in time, improves the accuracy of inversion, avoids the optimization from being trapped into non-physical or local minima and extends the time span for successful backward integrations. The influence of surface velocity can be seen in reconstructed thermal fields in which we find that the downwelling structures and the upper thermal boundary are better resolved than the upwellings and the lower thermal boundary. Essentially, the velocity data provides more constraints to shallower structure and this is evident from the sensitivity kernel of surface horizontal velocity to the underlying thermal field (Fig. S7, Supplementary Material). Nevertheless, the deeper structure is also substantially better constrained with surface velocities compared to inversions without as the deeper structure is strongly controlled by the shallow structure (e.g. downwellings control the locations of plume instability, E. Tan *et al.* 2002). Comparing reconstructions using inversion with and without surface velocity, we find that without velocity the locations of surface convergent are not accurate and the reconstructed initial state can only be recovered for a short period backwards in time ($t \sim -0.34t_{\rm tr}$); note that the fit to surface velocity is also poor if it is not used as a constraint (Fig. 5, Column 6).

Quantifying the correlation coefficient (see Supplementary Material) between reconstructed and ground truth thermal fields, we find that the error increases backwards in time with the joint inversion leading to the most robust recovery, followed by the reverse convection neural operator, the inversion by the terminal state and lastly the reverse buoyancy (Fig. 6a). This pattern generally holds most of the time. However, the performance of the reverse convection operator is better in some cases such as in time window 2 when smaller wavelength convection cells move laterally; with this time-dependence, the operator $\boldsymbol{F}_{\phi_7}^{-4}$ reconstructs the thermal state backwards over one transit time, and even outperforms the joint inversion approach in accuracy (Fig. S11, Supplementary Material).

Geophysical observations are noisy and so to explore the role of uncertainty on our four methods, we now compute thermal state reconstructions with the same methods but with observations polluted with 5 per cent pink noise (Fig. 7, Row 1, Columns 2 and 3). Compared with the previous results, large differences emerge in the reconstructions after adding the noise. The two direct methods, reverse buoyancy and the reverse convection neural operator fail to reconstruct even for short periods backward in time, although for different reasons. For the reverse buoyancy method, the pink noise in the terminal thermal state is interpreted as different short and long wavelengths features, in which the long wavelength ones especially cause substantial deviations in the flow upon reversal (Fig. 7, Column 2). As the reverse convection operator is an approximation to the ill-posed diffusion operator, it is quite sensitive to noise, which is quickly amplified during backwards integration, and the output becomes unstable within just a single step backward (Fig. 7, Column 3).

In contrast, the inverse approaches remain robust and can provide informative reconstructions even in the presence of noise when sufficient constraints are used in the time domain. When using only the terminal state as a constraint, realistic plume structures and some downwellings can be reconstructed to about $t = -0.34t_{\rm tr}$ (Fig. 7, Column 4)—further back than the direct methods—but still less stable in comparison to the ideal case without noise. Many thermal structures cannot be reconstructed further back than this. However, when the surface velocity data is added in a joint inversion, the thermal state can be reconstructed back to $t = -0.86t_{\rm tr}$ (Fig. 7, Column 5). Prior to this instant, the thermal structures become immersed in high-frequency variations resulting from the noised terminal state thermal field. Compared with the terminal thermal state produced by the joint inversion, we find without surface velocity constraints, the upwelling plumes in the upper 3/4 of the domain at terminal state cannot be fully recovered. In other words, the addition of the time-constraints helps to 'clean' the reconstruction from the structures that would be otherwise buried in noise. The correlations between reconstructions and ground truth in time verify the robustness of the joint inversion approach in the presence of noisy data (Fig. 6b).

## 4 DISCUSSION

We introduced three neural operators and examined their ability to learn aspects of mantle convection: (1) The Stokes neural operator $\boldsymbol{S}_\phi$ approximates the Stokes equations and maps temperature to velocity and pressure, (2) the forward convection neural operator $\boldsymbol{F}_\phi^{+n\Delta t}$ and (3) the reverse convection neural operator $\boldsymbol{F}_\phi^{-n\Delta t}$. The three neural operators have similar architecture, but differ in terms of prior information known of them, training strategy, well-posedness and the degree to which they can improve computational efficiency, especially for inverse problems. The convection neural operators were then used within and compared against different methods for thermal state reconstructions including inverse models that use different data types.

We started with the Stokes neural operator with prior knowledge from the underlying conservation equations. Trained by known physical laws, this physics-informed approach has the lowest pre-training expense with random inputs without invoking expensive numerical solves; in contrast, numerical solves of Stokes for training data set creation would contribute substantially to computational cost. The limitation of our physics informed approach is the difficulty of achieving the same precision for $\boldsymbol{S}_\phi$ as the data driven method (velocity channels of forward convection neural operators). Recall that our forward convection neural operators also output the Stokes velocities of the input thermal field, with a relative error less than 0.1 per cent. In contrast, the relative error from $\boldsymbol{S}_\phi$ is around 3–5 per cent. Such differences in prediction accuracy are also reflected in the power spectrum comparison between the neural operator and FEM outputs ($\boldsymbol{u}_x$ and $\boldsymbol{u}_z$ in Fig. S8 in Supplementary Material). At low wavenumbers, both $\boldsymbol{S}_\phi$ and $\boldsymbol{F}_{\phi_7}^{+1}$ closely match the FEM spectra. However, compared with $\boldsymbol{F}_{\phi_7}^{+1}$, $\boldsymbol{S}_\phi$ exhibits larger discrepancies at medium-higher wavenumbers, where it consistently overestimates the spectral power. While $\boldsymbol{F}_{\phi_7}^{+1}$ also shows a mild high-wavenumber excess—negligible for $\boldsymbol{u}_x$ and weak for $\boldsymbol{u}_z$, the overestimation is substantially more pronounced for $\boldsymbol{S}_\phi$, indicating that the physics-informed model introduces stronger spurious small-scale energy, particularly in the $\boldsymbol{u}_z$ channel. The discrepancy could be caused by a physics informed loss function which is stiff, as well as its gradients. In the physics informed approach, the loss function is composed of competing terms (eq. 23), while for the data driven one, the loss function is simpler, in form of $L = \|T_{\text{pred}} - T_{\text{true}}\|_2 + \|\boldsymbol{u}_{\text{pred}} - \boldsymbol{u}_{\text{true}}\|_2$. Lacking explicit supervision from data, the neural operator has to be self-guided towards the optimized state by the dynamically competing components in the loss, which is substantially more difficult for the optimization (S. Wang *et al.* 2022). We also reported those equation residuals on realistic convection Stokes outputs averaged over time window 1, of those derived by $\boldsymbol{S}_\phi$, as well as the data-driven $\boldsymbol{F}_\phi^{+n\Delta t}$ (Table B2). Though $\boldsymbol{S}_\phi$ is trained with a target of minimizing the PDE losses, $\boldsymbol{F}_\phi^{+n\Delta t}$ still arrives at a more accurate approximation of the Stokes operator: The equation residuals by $\boldsymbol{F}_\phi^{+n\Delta t}$ are in general one magnitude smaller than those by $\boldsymbol{S}_\phi$, consistent with the prediction errors reported in Tables 1 and 2. Thus, we suggest there exists a trade-off between pre-training cost and prediction precision.

The Stokes neural operator works for instantaneous geodynamic solves such as computing surface kinematics, dynamic topography, state of stress, etc. As those output quantities can be directly acquired, they can be used as constraints to invert for hidden states of the mantle. In addition, the Stokes equation can take in more mantle state variables including viscosity and chemical composition as inputs, as does the neural operator and lead to more realistic and comprehensive inversions. Along these lines using traditional methods, such as for an inversion of the thermochemical state of the mantle (A.M. Forte *et al.* 1994; A. Forte 2007) or for its nonlinear rheology (J. Hu *et al.* 2024) both from surface plate kinematics, Stokes neural operators could be an effective surrogate, speeding up such inversions. One limitation, however, for such instantaneous geodynamic inversions would be for models that involve the gravity field. The 5 per cent error in velocity leads to even higher errors for dynamic topography limiting accurate prediction of the geoid [which is a consequence of small differences between nearly equal, but opposite, contributions from driving buoyancy and boundary deflections (B.H. Hager *et al.* 1985) and requires topography to be computed at high accuracy]. Although the 5 per cent relative velocity prediction error of the neural Stokes is a moderate level of bias, it might cause a large deviation if called recursively to integrate a thermal system forward in time. The bias to the inversion result is contributed by an integration from both the model itself and the inaccuracies/uncertainties of observations, while the latter could be much larger compared with the 5 per cent model error. The model prediction error can be tolerated and the inversion could still remain informative with appropriate regularization. The Stokes neural operator might also be able to provide initial guesses for iterative numerical solvers on large scale, nonlinear systems.

In contrast, the forward convection neural operator is an example of data-driven operator discovery and maps between two thermal states separated by a long time interval, substantially greater than the CFL limit (potentially hundreds of times). Although we have some prior knowledge about this system and are able to solve the output from the input numerically, the two ends are not directly connected by a single set of PDEs. By transforming a process originally composed of multiple operators into a single neural one, computational efficiency is greatly improved. The consequence is that the model must be trained at substantial cost. Moreover, since thermal convection is an initial value problem, the forward operator must be called recursively, requiring a high level of accuracy for each step. Among the long-term convection patterns produced by forward convection neural operators trained with different time steps, a slight phase variation can be observed. While the effect is modest, it may reflect the fact that each model learns a distinct discrete-time evolution operator corresponding to its prescribed time interval. Different $\Delta t$ therefore defines different discrete flow maps, and small single-step approximation errors can propagate differently under recursive application, potentially leading to accumulated phase shifts over long horizons. This observation again highlights the trade-off between step size and error accumulation: Although a smaller $\Delta t$ allows the model to learn a more local and accurate single-step mapping, the increased number of recursive applications over a fixed physical time may still amplify small approximation errors. Our computations show that with sufficient training data, $\boldsymbol{F}_\phi^{+n\Delta t}$ achieves a high accuracy in modelling thermal convection for three to four transit times, roughly 150–200 Myr for mantle convection. To ensure the long-term prediction accuracy, the single step misfit must be maintained within a low level. With a given model architecture, we observed a strong dependence between training data size and the single step misfit within a finite size range (Fig. S3, Supplementary Material). In our case, a training data size greater than $10^4$ leads to 0.1 per cent single step prediction error, which is satisfactory for the following thermal state reconstruction problem. In practice, we adopted a database with around $10^{4.5}$ data pairs.

Although all forward convection neural operators trained under different Rayleigh numbers achieve satisfactory accuracy in temperature prediction, we observed that the prediction error moderately increases as Rayleigh numbers decrease. We suggest that this is likely an outcome that the spatial scales of temperature field increase as Rayleigh number decreases. The less sparse temperature field introduces a larger mis-predict error. Within the range tested, there is almost no indication that the long-term prediction error is increasing with time step size. However, we did not train an operator having an even larger step, because the step size ultimately determines the interval over which the inversion framework can incorporate chronological observations. Consequently, there is a balance between the ability to assimilate the constraints along the time axis and the number of recursive steps. If the step length is too large, the inversion will only have loose chronological constraints, while if too small, the computation will become more expensive. In the limit of large numbers of recursive integration steps, more GPU memory will be consumed as the entire computational chain is stored to enable auto-differentiation. In some cases with large memory usage, check points might have to be introduced during backward propagation. Hence, as a compromise we choose the neural operator with medium time step $\boldsymbol{F}_{\phi_7}^{+2}$ to carry out the inversion.

The ability of learning the mappings between bounded-energy function spaces, and in turn making zero-shot predictions across spatial resolutions as demonstrated by $\boldsymbol{S}_\phi$, is largely attributable to the inherent kernel architecture of neural operators. In our settings, this capability can be realized through data-driven training with sufficient Gaussian random field inputs which span a broad subset of admissible $L_2$ errors and thereby expose the model to diverse input realizations during training. A more critical challenge is zero-shot generalization across other dimensions of the problem class, including PDE parameters, geometry, characteristic scales and potentially even changes in the operator structure itself. Here we specifically examined parameter extrapolation governed by the Rayleigh number by training a variable Ra forward convection neural operator $\boldsymbol{F}^{+1}_{\phi_{\mathrm{Ra}}}$ and testing whether it can generalize beyond the training interval. Our results indicate that $\boldsymbol{F}^{+1}_{\phi_{\mathrm{Ra}}}$ can extend the learned Nu-Ra scaling trend to a limited neighbourhood outside the training range, suggesting some capacity for extrapolating continuous parameter-dependent behaviours. One shall note that Rayleigh number variation does not purely manifest a smooth transition in convection vigorousness: It can also induce discontinuous, regime changes in the underlying physics. The most fundamental example is the conduction–convection transition at the critical Rayleigh number $\mathrm{Ra}_{\mathrm{cr}}$, which lies below our training Ra range. In our results, $\boldsymbol{F}^{+1}_{\phi_{\mathrm{Ra}}}$ becomes unstable when evaluated at Rayleigh numbers a factor of two below the lower bound of the training interval, and thus would also fail to reproduce the transition behaviour across $\mathrm{Ra}_{\mathrm{cr}}$, an even lower Rayleigh number than the lower bound of our extrapolation range. In our case, we can not fully determine if the failed extrapolation is more likely a result of limited zero-shot capability, or the existence of a bifurcation outside the training Ra range. However, this observation might indicate the view that, for the current neural operator architectures (at least as instantiated in this study), zero-shot prediction, especially across *discontinuous* or *bifurcation-like* physics, remains limited, but based on the present results alone, we cannot rule out the possibility that neural operators under alternative architectures, training strategies or representations, could gain better extrapolation ability and in principle learn or anticipate such discontinuous behaviour. Understanding whether, and under what conditions, such transitions can be learned without explicit physical cues remains an open question for further investigations.

The direct neural operator model we put forward for mantle state reconstruction is the reverse convection neural operator. Analogous to the forward one, it is also an attempt of operator discovery from data, and can even be trained based on the identical data set prepared for the forward operator. Hence, for each forward surrogate model, a twin reversed operator can be trained. Our result shows that the reverse convection neural operator can indeed learn the ill-posed reverse process and predict backwards directly within a finite time interval, which is a breakthrough compared with numerical methods. However, it still shows some instabilities, including: (1) The performance could vary significantly dependent on the input thermal profile. We observed that if the convection pattern is dominated by four convection cells rather than two, the backwards integration by reverse convection neural operator remains stable for over one transit time. (2) It remains sensitive to noise and cannot be directly used in a realistic geophysical process with noisy observations. The most robust approach remains the inverse approach based on forward models. Nevertheless, if the input of the reverse operator lies precisely in the realistic function space, the backwards prediction is still feasible. After all, the significant efficiency of the direct formulation of a reverse operator remains appealing. Several strategies can be employed to improve the performance of a surrogate model for an ill-posed operator. For example, injecting noise into the input during training can enhance robustness and may be interpreted as a form of regularization (C.M. Bishop 1995). In addition, observational constraints, such as time-dependent surface kinematics, can be incorporated as additional input channels, leading to a conditional neural operator that learns mappings informed by auxiliary observations. This approach is commonly adopted in inverse problems and operator learning (e.g. A.M. Stuart 2010; L. Ardizzone *et al.* 2019; N. Kovachki *et al.* 2023), and conditioning can reduce ambiguity and improve empirical robustness by incorporating additional information. Nevertheless, neither of these strategies alters the intrinsic ill-posedness of the backward convection mapping. In this study, the reverse convection neural operator primarily serves to demonstrate its capability of learning an ill-posed mapping and the potential applications in mantle dynamics. A systematic investigation of noise-augmented or conditionally trained reverse operators is left for future work.

We also suggest that a denoise mapping from the observation to the true function space can be jointly used with the ill-posed neural operator. This pre-processing mapping on the observation can be parametrized by a neural network or formulated by a simpler inverse problem, and its outputs, might not only be denoised, but can also be transformed into other physical quantities if the observable one cannot be directly tackled by neural operators. For example, in mantle state reconstruction, the thermal state of mantle is indicated by seismic velocities using tomography. A transformation from noised and partially observed seismic velocities to a realistic thermal structure can be realized by an inversion constrained by the Stokes equation and wave equations. The pre-processing mapping could potentially make the usage of a reverse operator possible, even if we are using an inversion approach.

In addition to the learnability, accuracy and speedup of surrogate models, key is whether it is worthwhile to make a transition from traditional numerical solvers to surrogate ones, as the cost of training, especially the generation of training data, is considerable. For the purely physics informed Stokes neural operator, we have shown its learning ability without any pretraining cost, and its potential in instantaneous mantle dynamics modelling and inversions. For data-driven convection neural operators, although such surrogate models can speed up the computation substantially, their reliance on the training data set could cause a notable cost. To evaluate, one needs to consider the cost of the entire workflow. Compared with traditional ones, the neural operator workflow has the extra cost for training data preparation and training, in addition to forward and backward integrations. The optimized option is different depending on the application scenarios (evaluation details described in the Supplementary Material for the models with $\mathrm{Ra} = 10^7$): If the problem is a forward model integrated over a moderate interval (such as several transit times), the training of a neural operator might not be worthwhile; in this case, the cost of generating the training data will likely already exceed that of the traditional methods. For time-dependent inversions such as thermal state reconstruction, numerous iterations are required for optimization, with each iteration containing a forward and backward integration. The significant speedup by forward convection neural operators in integrations would manifest itself in such instances. Take the 2-D reconstruction across one transit time as an example, we suggest that the total cost of the neural operator workflow would be comparable with that of performing an inversion with traditional methods once. When multiple inversion trials are required, it will be far more efficient. In addition, as the computation speedup scales with problem size, the efficiency of neural-operator-based workflow would become more pronounced for larger scale problems (see Supplementary Material). The comparison indicates a strong prospect of neural operator's applications in time-dependent inverse problems.

In this study, we have focused on 2-D isoviscous thermal convection. Although simplified, this model captures key dynamical features relevant to mantle convection and serves as a demonstration of how neural operators can accelerate computation and reduce workflow costs. Nevertheless, such simplifications limit the range of geophysical problems that the current framework can address. To move toward, more realistic reconstructions of mantle evolution under plate tectonics, extensions to spherical-shell geometries and a more complex rheology would be required (e.g. P.J. Tackley 2000; T.W. Becker & L. Fuchs 2023). While they do not fundamentally alter the nature of the problem, such features would increase the cost of generating high-fidelity training data. Fortunately, recent developments in neural operator architectures are rapidly unfolding with advances showing even better performance than the FNO with the capability of learning complex physics at high-resolution within irregular domains (B. Bonev *et al.* 2023; Z. Li *et al.* 2023; Y. Shi *et al.* 2025). Although this study is restricted to a 2-D Cartesian domain with periodic lateral boundaries, the overall framework can, in principle, be extended to alternative geometries and boundary conditions through appropriate modifications to the PDE loss formulation (physics-informed) and data generation strategy (data-driven). In the current workflow, key modelling choices—including geometry, rheology and boundary conditions—are fixed throughout both training and deployment. Allowing greater solver-level flexibility would be desirable for accommodating diverse mantle dynamics scenarios without requiring complete retraining from scratch. In this context, some advanced training strategies, such as transfer learning, may allow a neural operator trained on a general mapping class to be efficiently fine-tuned for specific geometries, boundary configurations and even an alternative operator structure (e.g., M. McCabe *et al.* 2024). Continued advances in GPU hardware and large-scale training techniques are making the development and deployment of large neural operators increasingly feasible and efficient. Although computationally demanding, the construction of a fully 3-D neural operator for global convection appears within reach.

## 5 CONCLUSIONS

The utility of neural operators is demonstrated for 2-D, bottom-heated, Rayleigh–Bénard thermal convection. The results are consistent with earlier studies and have demonstrated that surrogate models in the form of neural operators can have a lower computational complexity when applied to both forward and backward calculations. The Fourier neural operator was used as the basic machine learning architecture. Starting by learning the Stokes system, an operator that can be expressed explicitly by a set of partial differential equations, the neural operator learns to solve the primitive variables from temperature with a purely physics informed approach without any training data, substantially reducing the pre-training cost. The physics informed Stokes neural operator can be potentially used as surrogate models for instantaneous geodynamics modelling and inversions. The mapping between two thermal states separated by temporal intervals significantly exceeding the CFL condition is approximated by forward convection neural operators through a data-driven training strategy. This surrogate model substantially accelerates forward convection computations through reducing the number of recursive steps. Furthermore, by exchanging the sequence of training data-pairs, the ill-posed reverse convection operators are approximated. The reverse convection operators are not unstable on integration as are the traditional numerical methods, and are able to predict previous mantle states directly. Several methods for thermal state reconstruction were developed and compared, including an inversion based on the forward convection neural operator and auto-differentiation. The results demonstrate this inverse approach is an accurate and robust method for reconstructing past states especially in the presence of observational noise. When chronological surface observations (surface kinematics) are added into a joint inversion along with the terminal thermal state, the method outperforms all other approaches. The accuracy of neural operators shows that they can serve as reliable surrogate models for mantle convection computations, tracking the well-benchmarked forward numerical integrations about three to four transit times, roughly 150–200 Myr for mantle convection, a timescale sufficient to resolve long-term tectonic process and the underlying dynamics. When embedded in the proposed joint-inversion framework for thermal state reconstruction, they recover the dominant thermal structures up to about one transit time earlier in our synthetic tests. We also found that for a reconstruction solve, the total cost of the neural-operator-based work flow, which is dominated by training, is approximately comparable to performing the inversion once using traditional numerical methods. The substantial speedup in both forward and reverse computations and their scaling with the system size shows the promise of global-scale mantle state reconstructions.

## ACKNOWLEDGMENTS

We thank Xiyuan Bao and David May for their careful and thoughtful comments which improved the manuscript. We acknowledge the use of the Anvil supercomputer at Purdue University, supported by the U.S. National Science Foundation (NSF), through ACCESS allocation EAR-160027.

## SUPPORTING INFORMATION

Supplementary data are available at *GJIRAS* online.

**Table S1.** Weighting coefficients of the objective function.
**Table S2.** Quantities used to evaluate the computational costs and the scaling of two methods.
**Table S3.** CPU (AMD EPYC™6954) and GPU (NVidia RTX™6000 Ada) times in seconds (s) for the numerical solver `Underworld` and forward convection neural operators by integrating the convection forward or propagate a gradient backward over one transit time.
**Figure S1.** Row 1: the source terms (input temperature $T$); Rows 2 and 3: solutions of $\boldsymbol{u}_x$ from `Underworld` and $\boldsymbol{S}_\phi$; Rows 4 and 5: solutions of $\boldsymbol{u}_z$ from `Underworld` and $\boldsymbol{S}_\phi$. The amplitudes of output velocity components are shown by numbers in each panel.
**Figure S2.** Prediction-error histograms for $\boldsymbol{F}^{+1}_{\phi_7}$, $\boldsymbol{F}^{+2}_{\phi_7}$ and $\boldsymbol{F}^{+4}_{\phi_7}$. In each subplot, the dashed vertical lines mark the 10th, median and 90th percentiles; the $x$-axis ticks indicate the minimum, mean and maximum errors, respectively.
**Figure S3.** Forward convection neural operator performance under different training data sizes.
**Figure S4.** (a) Typical training data pairs used to train $\boldsymbol{F}^{+1}_{\phi_{\mathrm{Ra}}}$ near Ra $= 10^{6.5}$, randomly sampled over the training data integration time scope. (b) Steady convection produced by $\boldsymbol{F}^{+1}_{\phi_{\mathrm{Ra}}}$.
**Figure S5.** Comparisons between forward computations using $\boldsymbol{F}^{+n}_{\phi_6}$ and `Underworld`, Ra $= 10^6$. Rows 1 to 4 show the forward evolution snapshots of the systems. Column 1: temperature snapshots and velocity streamlines computed by $\boldsymbol{F}^{+1}_{\phi_6}$, system integrated by $\boldsymbol{F}^{+1}_{\phi_6}$; Column 2: temperature and velocity computed by `Underworld`, system integrated by `Underworld`; Column 3: velocity computed by $\mathcal{S}_\phi$ based on thermal fields integrated by `Underworld` as inputs. Rows 5 and 6: Nu and $\boldsymbol{u}_{x0}$ tracked in systems integrated with different methods. Black short bars indicate the instants where the snapshots within rows 1 to 4 are chosen.
**Figure S6.** Comparisons between forward computations using $\boldsymbol{F}^{+n}_{\phi_5}$ and `Underworld`, Ra $= 10^5$. Rows 1 to 4 show the forward evolution snapshots of the systems. Column 1: temperature snapshots and velocity streamlines computed by $\boldsymbol{F}^{+1}_{\phi_5}$, system integrated by $\boldsymbol{F}^{+1}_{\phi_5}$; Column 2: temperature and velocity computed by `Underworld`, system integrated by `Underworld`; Column 3: velocity computed by $\mathcal{S}_\phi$ based on thermal fields integrated by `Underworld` as inputs. Rows 5 and 6: Nu and $\boldsymbol{u}_{x0}$ tracked in systems integrated with different methods. Black short bars indicate the instants where the snapshots within rows 1 to 4 are chosen.
**Figure S7.** Green's functions of surface velocity to the thermal structure $\tilde{V}_k(z)$, shown by perturbation structures' wave numbers $k$ in columns. Vertical axis $z$ is the perturbation structures' burial depth and the horizontal axis denotes the amplitudes of the kernel. Neural operators' solutions from $\boldsymbol{F}^{+1}_{\phi_7}$ and $\boldsymbol{S}_\phi$ are compared against the analytical solution (eq. S8).
**Figure S8.** Time-averaged directional power spectra of the predicted fields. Shown are $T$, $\boldsymbol{u}_x$ and $\boldsymbol{u}_z$ predicted by $\boldsymbol{F}^{+1}_\phi$, and $\boldsymbol{u}_x$ and $\boldsymbol{u}_z$ predicted by $\mathbf{S}_\phi$, all compared against the FEM reference solution. The horizontal axis is the (non-negative) Fourier mode index $m$; physical wavenumbers can be recovered via $k_x = m_x/L_x$ and $k_z = m_z/D$. The zero mode is excluded and spectra are plotted for $m \geq 1$.
**Figure S9.** Optimization curves of the joint inversions, showing the objective-function values (eq. 13) and the misfit values for the two observations. (a) Joint inversion. (b) Joint inversion with polluted synthesized observations. (c) Joint inversion using stochastic gradient descent. The triangles in panels (a) and (b) denote the 25 000th optimization iteration, at which the optimization has converged and the results are used as demonstrations in Fig. 5.
**Figure S10.** Reconstruction performances of different methods over time window 2 with clean observations. Reversal time steps shown by row (from top to bottom backwards in time). Column 1: a ground-truth thermal convection sequence computed by `Underworld` forward in time from bottom to top; Columns 2–5: reconstruction with: reverse buoyancy; reverse convection operator; inversion with terminal state only; joint inversion; Column 6: At each time step, the surface horizontal velocity profiles compare the clean observations with predictions from the inversion with terminal state only and the joint inversion.
**Figure S11.** Correlation coefficients of reconstructed thermal fields with ground-truth fields versus backward time within time window 2. Each line represents a different reconstruction method. (a) Reconstruction with synthesized observations (no noise). (b) Reconstruction with synthesized observations polluted with 5 per cent pink noise.

Please note: Oxford University Press is not responsible for the content or functionality of any supporting materials supplied by the authors. Any queries (other than missing material) should be directed to the corresponding author for the paper.

## DATA AVAILABILITY

The code and data used in this study are publicly available through the GitHub repository at https://github.com/kong-chenxi/mantle-convection-neural-operators.

## APPENDIX A: NEURAL OPERATOR ARCHITECTURES

The detailed neural operator hyperparameters are listed in Table A1 .

**Table A1.** The architectural hyperparameters of neural operators trained in this study.

| Neural operators | Modes[a] | Hidden Channels[a] | Number of FB[c] | Lifting ratio[b] | Projection ratio[b] | Rank[c] |
|---|---|---|---|---|---|---|
| $\boldsymbol{F}_{\phi_5}^{+n\Delta t}$ | 65 | 128 | 6 | 2 | 2 | 0.25 |
| $\boldsymbol{F}_{\phi_6}^{+n\Delta t}$ | 129 | 128 | 6 | 2 | 2 | 0.1 |
| $\boldsymbol{F}_{\phi_7}^{\pm n\Delta t}$ | 129 | 128 | 6 | 2 | 2 | 0.1 |
| $\boldsymbol{F}_{\phi_{\mathrm{Ra}}}^{+\Delta t}$ | 129 | 128 | 6 | 2 | 2 | 0.1 |
| $\boldsymbol{S}_{\phi}$ | 65 | 128 | 5 | 1 | 1 | 0.25 |

[a]FB hyperparameters.
[b]MLP hyperparameters.
[c]TFNO hyperparameters.

## APPENDIX B: PDE LOSSES OF PHYSICS INFORMED STOKES NEURAL OPERATOR

PDE losses of the physics informed Stokes neural operator are evaluated by applying FD operators to the operator outputs, namely $\boldsymbol{u}$ and $p$. To ensure numerical stability of the discrete Stokes operator, we adopted a staggered grid formulation commonly used in FD geodynamic solvers (T. Gerya 2019). Specifically, $p$ is evaluated at cell centres, $\boldsymbol{u}_x$ is evaluated at the centres of cell faces normal to the horizontal direction and $\boldsymbol{u}_z$ is evaluated at the centres of cell faces normal to the vertical direction (Fig. B1). With this staggered arrangement, the continuity loss $L_{\mathrm{C}}$ is evaluated at cell centres, the horizontal momentum loss $L_{\mathrm{Mx}}$ at the locations of $\boldsymbol{u}_x$ and the vertical momentum loss $L_{\mathrm{Mz}}$ at the locations of $\boldsymbol{u}_z$. We refer to these locations as the query points of the PDE losses. The specific formulation of losses $L_{\mathrm{C}}$, $L_{\mathrm{Mx}}$ and $L_{\mathrm{Mz}}$, and the distributions of the PDE loss query points can be found in eqs (B1)–(B3).

Fig. S1 (Supplementary Material) shows the velocity predictions from $\boldsymbol{S}_\phi$ with inputs from convection patterns under different Rayleigh numbers ($10^5$, $10^6$, $10^7$), and different resolutions ($65 \times 65$, $129 \times 129$, $257 \times 257$), respectively. Table B1 lists the weighting coefficients used to train $\boldsymbol{S}_\phi$.

The continuity loss $L_{\mathrm{C}}$ evaluated on the cell centres is:

$$L_{\mathrm{C}} = \left\| \frac{\partial(\boldsymbol{u})_x}{\partial x} + \frac{\partial(\boldsymbol{u})_z}{\partial z} \right\|_{\mathrm{F}} = \left[ \sum_{i=1}^{N_x-1} \sum_{j=1}^{N_z-1} \left( \frac{\boldsymbol{u}_{x,i+1,j} - \boldsymbol{u}_{x,i,j}}{\Delta x} + \frac{\boldsymbol{u}_{z,i,j+1} - \boldsymbol{u}_{z,i,j}}{\Delta z} \right)^2 \right]^{1/2}. \quad \text{(B1)}$$

The momentum losses $L_{\mathrm{Mx}}$ and $L_{\mathrm{Mz}}$ are evaluated on locations of $\boldsymbol{u}_x$ and $\boldsymbol{u}_z$, respectively:

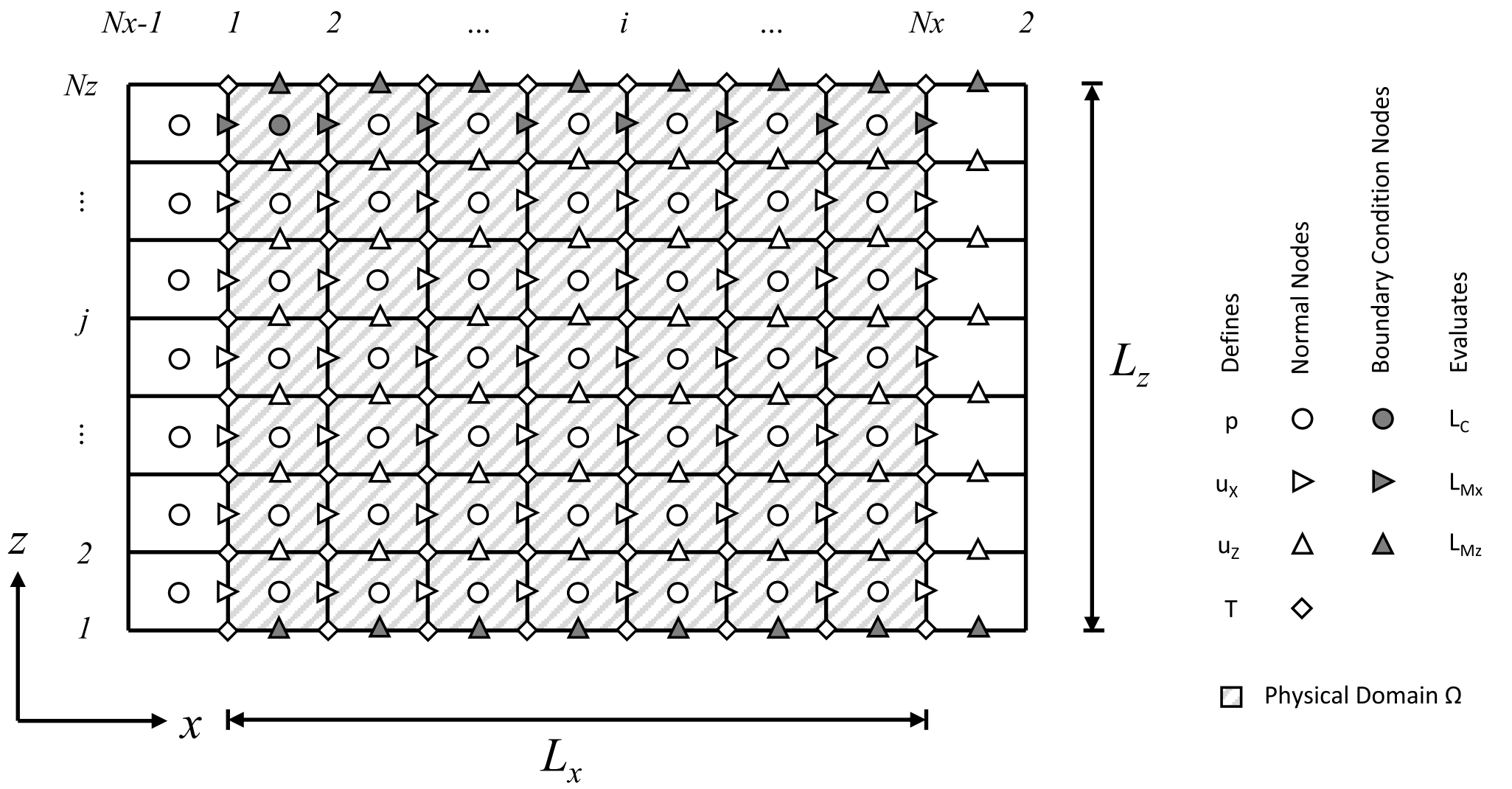

**Figure B1.** The evaluation method of PDE losses for the training of $\boldsymbol{S}_\phi$. Stencil of the staggered grid used to evaluate the PDE loss of the physics informed Stokes neural operator modified from T. Gerya (2019). The grey shaded area shows the physical domain Ω. Note that as the two side boundaries of Ω are periodic, the two extra columns of subdomains attached to the sides of Ω are physically identical.

**Table B1.** Weighting coefficients of PDE losses to train the physics informed Stokes neural operator.

| Training phase | Input resolution | $\beta_{\mathrm{C}}$ | $\beta_{\mathrm{Mx}}$ | $\beta_{\mathrm{Mz}}$ | $\beta_{\mathrm{u}}$ | $\beta_{\mathrm{p}}$ | $\beta_{\mathrm{BF}}$ | $\beta_{\mathrm{BP}}$ | $\beta_{\mathrm{G}}$ |
|---|---|---|---|---|---|---|---|---|---|
| Stage I | $129 \times 129$ | 128 | 1 | 1 | $10^{-3.5}$ | $10^{2}$ | 128 | 128 | 128 |
| Stage II | $129 \times 129$ | 256 | 2 | 3 | $10^{-3.5}$ | $10^{2}$ | $128^{2}$ | $128^{2}$ | 12.8 |

**Table B2.** Physical residuals of the Stokes outputs.

| Neural operators | $L_{\mathrm{C}}$ | $L_{\mathrm{Mx}}$ | $L_{\mathrm{Mz}}$ | $L_{\mathrm{BF}}$ | $L_{\mathrm{BP}}$ |
|---|---|---|---|---|---|
| $\boldsymbol{F}_{\phi_7}^{+1}$ | 0.1024 | 77.3069 | 720.5225 | 0.1664 | 0.0502 |
| $\boldsymbol{S}_{\phi}$ | 1.3791 | 579.6528 | 1361.8393 | 1.2458 | 3.3424 |

Since $\boldsymbol{F}_{\phi_7}^{+1}$ does not output pressure, when calculating its momentum residuals, we used pressure evaluation from $\boldsymbol{S}_{\phi}$. All the residuals are averaged by the number of query points. The physical residuals quantify the inconsistency of the predicted fields with the dimensionless conservation equations. After normalization by the domain size, the continuity residual $L_{\mathrm{C}}$ provides an average measure of the local dimensionless volume flux imbalance associated with violations of incompressibility. Similarly, after accounting for the velocity-output scaling used in the loss definition and the domain size normalization, $L_{\mathrm{Mx}}$ and $L_{\mathrm{Mz}}$ measure the residual unbalanced dimensionless body force density in the horizontal and vertical momentum balances, respectively.

$$
\begin{aligned}
L_{\mathrm{Mx}} =& \left\| \frac{\partial^2 \boldsymbol{u}_x}{\partial x^2} + \frac{\partial^2 \boldsymbol{u}_x}{\partial z^2} - \beta_{\mathrm{p}} \frac{\partial p}{\partial x} \right\|_{\mathrm{F}} \\
=& \left[ \sum_{i=1}^{N_x-2} \sum_{j=1}^{N_z-3} \left( \frac{\boldsymbol{u}_{x,i+2,j+1} + \boldsymbol{u}_{x,i,j+1} - 2\boldsymbol{u}_{x,i+1,j+1}}{\Delta x^2} + \frac{\boldsymbol{u}_{x,i+1,j+2} + \boldsymbol{u}_{x,i+1,j} - 2\boldsymbol{u}_{x,i+1,j+1}}{\Delta z^2} \right.\right. \\
& \left. - \beta_{\mathrm{p}} \frac{p_{i+1,j+1} - p_{i,j+1}}{\Delta x} \right)^2 \\
& + \sum_{j=1}^{N_z-3} \left( \frac{\boldsymbol{u}_{x,2,j+1} + \boldsymbol{u}_{x,N_x-1,j+1} - 2\boldsymbol{u}_{x,1,j+1}}{\Delta x^2} + \frac{\boldsymbol{u}_{x,N_x,j+2} + \boldsymbol{u}_{x,N_x,j} - 2\boldsymbol{u}_{x,N_x,j+1}}{\Delta z^2} \right. \\
& \left.\left. - \beta_{\mathrm{p}} \frac{p_{1,j+1} - p_{N_x-1,j+1}}{\Delta x} \right)^2 \right]^{1/2},
\end{aligned} \tag{B2}
$$

$$
\begin{aligned}
L_{\mathrm{Mz}} =& \left\| \frac{\partial^2 \boldsymbol{u}_z}{\partial x^2} + \frac{\partial^2 \boldsymbol{u}_z}{\partial z^2} - \beta_{\mathrm{p}} \frac{\partial p}{\partial z} + \beta_{\mathrm{u}} \mathrm{Ra}(T - T_{\mathrm{c}}) \right\|_F \\
=& \left[ \sum_{i=1}^{N_x-3} \sum_{j=1}^{N_z-2} \left( \frac{\boldsymbol{u}_{z,i+2,j+1} + \boldsymbol{u}_{z,i,j+1} - 2\boldsymbol{u}_{z,i+1,j+1}}{\Delta x^2} + \frac{\boldsymbol{u}_{z,i+1,j+2} + \boldsymbol{u}_{z,i+1,j} - 2\boldsymbol{u}_{z,i+1,j+1}}{\Delta z^2} \right.\right. \\
& \left. - \beta_{\mathrm{p}} \frac{p_{i+1,j+1} - p_{i+1,j}}{\Delta z} + \beta_{\mathrm{u}} \mathrm{Ra} \left( \frac{T_{i+1,j+1} + T_{i+2,j+1}}{2} - T_{\mathrm{c}} \right) \right)^2 \\
& + \sum_{j=1}^{N_z-2} \left( \frac{\boldsymbol{u}_{z,2,j+1} + \boldsymbol{u}_{z,N_x-1,j+1} - 2\boldsymbol{u}_{z,1,j+1}}{\Delta x^2} + \frac{\boldsymbol{u}_{z,1,j+2} + \boldsymbol{u}_{z,1,j} - 2\boldsymbol{u}_{z,1,j+1}}{\Delta z^2} \right. \\
& \left. - \beta_{\mathrm{p}} \frac{p_{1,j+1} - p_{1,j}}{\Delta z} + \beta_{\mathrm{u}} \mathrm{Ra} \left( \frac{T_{1,j+1} + T_{2,j+1}}{2} - T_{\mathrm{c}} \right) \right)^2 \\
& + \sum_{j=1}^{N_z-2} \left( \frac{\boldsymbol{u}_{z,1,j+1} + \boldsymbol{u}_{z,N_x-2,j+1} - 2\boldsymbol{u}_{z,N_x-1,j+1}}{\Delta x^2} + \frac{\boldsymbol{u}_{z,N_x-1,j+2} + \boldsymbol{u}_{z,N_x-1,j} - 2\boldsymbol{u}_{z,N_x-1,j+1}}{\Delta z^2} \right. \\
& \left.\left. - \beta_{\mathrm{p}} \frac{p_{N_x-1,j+1} - p_{N_x-1,j}}{\Delta z} + \beta_{\mathrm{u}} \mathrm{Ra} \left( \frac{T_{N_x-1,j+1} + T_{N_x,j+1}}{2} - T_{\mathrm{c}} \right) \right)^2 \right]^{1/2},
\end{aligned} \tag{B3}
$$

and,

$$
L_{\mathrm{M}} = L_{\mathrm{Mx}} + L_{\mathrm{Mz}}. \tag{B4}
$$

**Table C1.** CFL time steps of steady convection patterns at different Rayleigh numbers.

| Ra | $10^5$ | $10^6$ | $10^7$ |
|---|---|---|---|
| CFL time step | $2.12 \times 10^{-5}$ | $2.27 \times 10^{-6}$ | $2.5 \times 10^{-7}$ |

**Table C2.** Benchmark test on the steady convection state generated by forward convection neural operators. $\mathrm{Nu_s}$: surface Nusselt number; $v_{\mathrm{rms}}$: non-dimensional rms velocity; $q_i$: non-dimensional temperature gradients at the corners of the cell.

| Model | Grid | $\mathrm{Nu_s}$ | $v_{\mathrm{rms}}$ | $q_1$ | $q_2$ |
|---|---|---|---|---|---|
| | | | $\mathrm{Ra} = 10^5$ | | |
| $\boldsymbol{F}_{\phi_5}^{+1}$ | $64 \times 64$ | 10.2849 | 195.1245 | 18.2477 | 0.6971 |
| REF | ext. | 10.5341 | 193.2145 | 19.0795 | 0.7228 |
| | | | $\mathrm{Ra} = 10^6$ | | |
| $\boldsymbol{F}_{\phi_6}^{+1}$ | $128 \times 128$ | 21.2338 | 835.2131 | 42.8754 | 0.8139 |
| REF | ext. | 21.9725 | 833.9897 | 45.9642 | 0.9443 |

## APPENDIX C: DATA-DRIVEN CONVECTION NEURAL OPERATORS

The integration time steps of data-driven convection neural operators are significantly larger than the CFL time steps with regards to convection patterns at specific Rayleigh numbers.

We tracked two variables: Mid depth Nusselt number Nu, maximum horizontal velocity of the convection cell $\boldsymbol{u}_{x0}$, of convection over time window 1 computed by the forward convection neural operator and `Underworld` to evaluate the long-term prediction stability and accuracy. Nu is given by

$$\mathrm{Nu} = \left. \frac{< \boldsymbol{u}_z T > - < \partial_z T >}{< \partial_z T >} \right|_{z=D/2} \tag{C1}$$

and $\boldsymbol{u}_{x0}$ is calculated by searching for the largest velocity components in the domain.

The long-term evolution is measured in units of the transit time $t_{\mathrm{tr}}$, defined as the characteristic time required for a fluid parcel to traverse the vertical extent of the domain. In this study, $t_{\mathrm{tr}}$ is estimated as the average traveltime of material parcels along the vertical cores of the upwelling and downwelling regions. Specifically, we compute the traveltime along representative streamline segments in the hot upwelling and cold downwelling cores, and then average over a sufficiently long time interval $t_0$ during the steady convection state:

$$t_{\mathrm{tr}} = \frac{1}{2t_0} \int_0^{t_0} \left[ \int_{\mathrm{upwelling}} \frac{\mathrm{d}l}{|\boldsymbol{u}|} + \int_{\mathrm{downwelling}} \frac{\mathrm{d}l}{|\boldsymbol{u}|} \right] \mathrm{d}t. \tag{C2}$$

In addition to the mean errors reported in Table 2, Fig. S2 (Supplementary Material) shows histograms of the pointwise prediction-error distributions over the selected time window.

Besides the example of $\mathrm{Ra} = 10^7$ provided in the main text, we also present the examples of convection modelling by forward convection neural operators with lower Rayleigh numbers, Ra= $10^6$ and $10^5$ (Figs S5 and S6, Supplementary Material).

Under a data-driven approach, we also explored how the training data size could affect the learning outcome. Fig. S3 (Supplementary Material) shows the test error of forward convection neural operator $\boldsymbol{F}_{\phi_7}^{+1}$ versus the number of training data pairs. Single step prediction test errors of three output channels averaged over time window 1 are measured.

Fig. S4(a) (Supplementary Material) shows the typical training data pairs used in the training data integration time scope near $\mathrm{Ra} = 10^{6.5}$, a stage that is still distant from the steady convection state; Fig. S4(b) (Supplementary Material) shows the steady convection state produced by the trained $\boldsymbol{F}_{\phi_{\mathrm{Ra}}}^{+1}$ though a long forward time integration. The training time is roughly proportional to the training set size in our practice. For the $\sim 10^{4.5}$ training sample pairs case, the training GPU time is reported in the Supplementary Material.

In order to test the validity of our approach, we also tested our data-driven neural operators against the benchmark test for steady mantle convection with constant viscosity in a square box, the case 1 in B. Blankenbach *et al.* (1989), with two Rayleigh numbers $10^5$ and $10^6$. The dimensionless temperature is fixed to 1 at the bottom and 0 at the top. All boundaries are free-slip, with computational domain aspect ratio 1/1. The reference values are taken from B. Blankenbach *et al.* (1989), which are based on extrapolated infinite-resolution results of the finite element code by U. Christensen (1984). The results are listed in Table C2. All quantities are defined following B. Blankenbach *et al.* (1989).